# Heterogeneous Exposures to Systematic and Idiosyncratic Risk across Crypto Assets: A Divide-and-Conquer Approach[★]

Nektarios Aslanidis[a], Aurelio F. Bariviera[a], George Kapetanios[b], Vasilis Sarafidis[c]

[a]*Universitat Rovira i Virgili ECO-SOS*
[b]*King's College London*
[c]*Brunel University London*

**Abstract**

This paper analyzes realized return behavior across a broad set of crypto assets by estimating heterogeneous exposures to idiosyncratic and systematic risk. A key challenge arises from the latent nature of broader economy-wide risk sources: macro-financial proxies are unavailable at high-frequencies, while the abundance of low-frequency candidates offers limited guidance on empirical relevance. To address this, we develop a two-stage "divide-and-conquer" approach. The first stage estimates exposures to high-frequency idiosyncratic and market risk only, using asset-level IV regressions. The second stage identifies latent economy-wide factors by extracting the leading principal component from the model residuals and mapping it to lower-frequency macro-financial uncertainty and sentiment-based indicators via high-dimensional variable selection. Structured patterns of heterogeneity in exposures are uncovered using Mean Group estimators across asset categories. The method is applied to a broad sample of crypto assets, covering more than 80% of total market capitalization. We document short-term mean reversion and significant average exposures to idiosyncratic volatility and illiquidity. Green and DeFi assets are, on average, more exposed to market-level and economy-wide risk than their non-Green and non-DeFi counterparts. By contrast, stablecoins are less exposed to idiosyncratic, market-level, and economy-wide risk factors relative to non-stablecoins. At a conceptual level, our study develops a coherent framework for isolating distinct layers of risk in crypto markets. Empirically, it sheds light on how return sensitivities vary across digital asset categories – insights that are important for both portfolio design and regulatory oversight.

[★]Acknowledgements: The authors would like to thank L. Coroneo, T. Diasakos, T. Panagiotidis, D. Robertson, C. Savva, A. Urquhart, as well as participants at the 29th International Panel Data Conference (Orléans, 4-5 July, 2024), and the Macro-Finance Cluster Seminar of the University of York (22 October, 2024) for useful comments and suggestions. Aslanidis acknowledges financial support by the Spanish Government, Ministry of Science and Innovation under the Project reference PID2022-137382NB-I00.

*Email addresses:* `nektarios.aslanidis@urv.cat` (Nektarios Aslanidis),

# 1. Introduction

This paper analyzes realized return behavior across crypto assets by estimating heterogeneous exposures to idiosyncratic and systematic risk, including both market-level and economy-wide components. Our objectives are threefold. First, we assess the contribution of idiosyncratic risk to realized return behavior, both on its own and relative to core market-level risk components. Second, we evaluate the extent to which crypto assets are becoming more integrated with, or insulated from, traditional asset classes such as equities, bonds, and gold. Third, we investigate – and provide evidence for – distinct and meaningful differences in return exposures to both market-level and economy-wide risk across different categories of crypto assets.[1]

The market of crypto assets has evolved from a fringe phenomenon to a globally significant asset class, with a market capitalization exceeding $3.4 trillion and daily trading volumes of $125 billion (Coinmarket, 2025). Institutional adoption is rising, with major players like BlackRock offering retail-facing crypto products (Hur, 2024). Meanwhile, regulatory clarity is advancing through frameworks such as the EU's Markets in Crypto Assets (MiCA) regulation and increased oversight from U.S. agencies including the SEC, CFTC, FinCEN, and FTC. These developments, alongside ongoing innovation in blockchain infrastructure, are reshaping both investor behavior and the broader financial ecosystem (Cong et al., 2023a; Rzayev et al., 2024).

Despite recent advances, the digital asset market remains structurally distinct from traditional financial markets. It is characterized by persistent inefficiencies (Urquhart, 2016; Bariviera, 2017), limited diversification opportunities (Corbet et al., 2018; Anyfantaki et al., 2021), heterogeneous trading microstructures (Panagiotidis et al., 2019), and security vulnerabilities such as ransomware attacks (Cong et al., 2025).

Moreover, the market retains a strong speculative character, partly because fundamentals play only a limited role in shaping return dynamics. Instead, asset valuations are often driven by broader macro-financial forces, such as policy uncertainty, geopolitical risk, shifts in business and investor sentiment, and narrative-driven expectations. We refer to these aggregate influences collectively as "economy-wide risk". As a result, the market is especially prone to bubbles (Hafner, 2020; Liu and Tsyvinski, 2021; Lambrecht et al., 2025), price manipulation (Cong et al., 2023b), and episodes of extreme instability.

A key empirical challenge is that economy-wide risk is inherently latent: macro-financial proxies are typically measured using indicators available only at monthly or quarterly fre-

aurelio.fernandez@urv.cat (Aurelio F. Bariviera), george.kapetanios@kcl.ac.uk (George Kapetanios), vasilis.sarafidis@brunel.ac.uk (Vasilis Sarafidis) quencies, with almost no data recorded at high frequency.[2] In addition, there exists an abundance of low-frequency candidate proxies for economy-wide risk – spanning uncertainty, sentiment, macroeconomic performance, and policy conditions – but no consensus on which are most relevant for digital assets. This phenomenon of factor proliferation, commonly referred to as the "factor zoo"(Cochrane, 2011; Harvey et al., 2016; Hou et al., 2020), can be especially acute in the context of crypto assets, where empirical research remains nascent and the financial ecosystem continues to evolve rapidly.

[1] We use the term "crypto assets" to refer collectively to both native cryptocurrencies and tokens. Native cryptocurrencies are the intrinsic assets of a given blockchain and are used to secure and operate the network. Examples include BTC on the Bitcoin blockchain and ETH on Ethereum. Their issuance is typically governed by the protocol itself, either through a predetermined schedule or as compensation for network validators (e.g., mining or staking rewards). On the other hand, tokens are digital assets issued via smart contracts on top of existing blockchains. These do not have independent ledger infrastructure. For example, PAXG and USDC are ERC-20 tokens operating on Ethereum. Tokens can represent a wide range of functionalities, from utility and governance rights to representations of real-world assets. Their creation, distribution, and rules of operation are encoded in smart contracts on the host blockchain. We consider both native cryptocurrencies and tokens, spanning a range of blockchain platforms and use cases.

This paper develops a novel two-stage "divide-and-conquer" modeling framework designed to address several of the empirical challenges outlined above. The two stages arise from partitioning systematic risk into two groups: market-level risk – observed at high frequency – and economy-wide risk, which reflects macro-financial shocks proxied at lower frequency. In Stage 1, we estimate heterogeneous exposures to idiosyncratic and market-level risk only, using asset-specific instrumental variable (IV) regressions. Despite the endogeneity of idiosyncratic risk arising from confounding effects of latent economy-wide influences, the IV estimator is shown to yield consistent estimates of the heterogeneous exposures. This is achieved by exploiting the fact that economy-wide risk exhibits a common factor structure, affecting all assets albeit with varying intensity. This structure enables the construction of valid instruments through the extraction of shared variation from the model covariates via principal components analysis (PCA), thereby purging the influence of the latent common component.

In Stage 2, we shift focus to the estimation of exposures to economy-wide risk. To address its latent nature – stemming from the lower frequency and high dimensionality of available proxies – our approach proceeds as follows. First, we extract the leading principal component from the model residuals using PCA. We then average the weekly values of this leading component within each calendar month to align its frequency with that of the candidate economy-wide indices. Next, we project this monthly-aggregated series onto a broad set of macro-financial, policy-related, and sentiment-based indicators that serve as proxies for economy-wide risk. Relevant predictors are selected using the multiple testing boosting (MTB) method recently introduced by Kapetanios et al. (2025), designed for parsimonious variable selection in high-dimensional settings. This combined PCA-MTB approach is shown to perform well in finite samples, with supporting evidence provided in Appendix C. We then estimate asset-specific exposures by regressing each asset's residual variation on the selected economy-wide risk proxies.

As a final step, we apply a Mean Group estimator to the asset-level exposures from both stages to summarize cross-sectional variation and uncover structured patterns of heterogeneity across different categories of crypto assets.

We implement our divide-and-conquer approach to a broad sample of crypto assets that collectively account for more than 80% of the total crypto market capitalization, observed weekly from January 2020 to December 2022. Idiosyncratic risk is proxied by volatility, illiquidity, and lagged returns, reflecting frictions and short-term return predictability. Market-level risk is proxied by market returns and volatility from four major asset classes: crypto, equities, bonds, and gold. In addition, we include size and momentum factors specific to the crypto market, which have been shown to explain cross-sectional return variation (Liu et al., 2022). To represent economy-wide influences, we compile 35 monthly indicators covering uncertainty, sentiment, market attention, and global financial conditions (listed in Appendix A) intended to capture diverse dimensions of macro-financial risk. These proxies are drawn from several influential studies and – so far as we are aware – this is the first to integrate such a diverse collection of indicators into a unified modeling framework.

---

[2] Even if crypto asset data can be meaningfully harmonized to a lower frequency, discarding high-frequency variation in returns may weaken identification of asset-specific exposures to idiosyncratic and market level risk components.

The empirical results are organized around two main areas of contribution: estimation of exposures to idiosyncratic and systematic risk – both at the asset level and aggregated across the full sample; and analysis of patterns of heterogeneity across three key subgroups: green vs non-green assets, stablecoins vs non-stablecoins, and DeFi vs non-DeFi assets.

The comparative analysis across subgroups uncovers clear segmentation in how different types of crypto assets respond to distinct sources of risk. On average, green assets are more sensitive to both market-level and economy-wide risk than their non-green counterparts. Moreover, they exhibit a pronounced response to climate policy uncertainty, consistent with investor preferences and climate-aligned narratives (Pastor et al., 2021, 2020). To illustrate this contrast, we conduct a stylized experiment examining return responses to a 10% increase in the mean of four key variables: idiosyncratic volatility and illiquidity, crypto market volatility and size. For green digital assets, approximately 61% of the variation in total returns can be attributed to market-level factors, while only 39% is due to idiosyncratic sources. In contrast, over 80% of the variation in non-green crypto assets is driven by idiosyncratic channels.

Stablecoins exhibit strong mean reversion and negligible average sensitivity to idiosyncratic volatility, and remain relatively insulated from broader economy-wide risk, consistent with their design as low-volatility, peg-maintained crypto assets. By contrast, non-stablecoins show greater exposure to both idiosyncratic and market-level risk factors.

DeFi assets are more sensitive to systematic risk, particularly market size and global uncertainty, reflecting their dependence on platform scale and liquidity concentration. In contrast, non-DeFi assets respond more to momentum and gold market movements, consistent with greater exposure to short-term trends and safe-haven flows.

These findings contribute to a growing literature that seeks to model the unique dynamics of crypto assets. For example, prior work by Liu et al. (2022) adapts the Fama-French framework to show that size and momentum help explain crypto asset returns, while Cong et al. (2022) extend this approach by introducing value and network adoption premia. Cong et al. (2021) and Sockin and Xiong (2023) emphasize the role of investor attention and information frictions, while Augustin et al. (2023) and Gordon et al. (2023) highlight structural features such as segmentation and noise trading. We build on these insights and show that idiosyncratic risk – particularly volatility and illiquidity – is, on average, positively associated with realized returns. This suggests that asset-specific frictions remain a fundamental source of return variation, consistent with incomplete diversification and execution risk. Our results further point to growing integration with traditional markets, as crypto assets exhibit significant average exposures to equity returns and volatility, indicating increasing co-movement with stock conditions. This marks a shift from earlier views of crypto market isolation (Corbet et al., 2018; Aslanidis et al., 2019).

The results of our study carry several important implications. For investors, the distinct patterns of return sensitivity identified across asset categories may serve as a key determinant of portfolio allocation and risk management. For policymakers, the strong sensitivity of crypto returns to traditional asset classes suggests that macroeconomic and regulatory developments may have significant spillover effects. Moreover, climate-conscious investment narratives appear to amplify exposure to policy risk for green digital assets. In parallel, the distinct behavior of stablecoins raises questions similar to those surrounding Central Bank Digital Currencies, particularly regarding financial stability and monetary transmission mechanisms (Ferrari Minesso et al., 2022).

Beyond its application to digital asset markets, our divide-and-conquer approach is broadly applicable, even in settings where idiosyncratic risk is minimal but the risk environment is high-dimensional and characterized by factors measured at varying frequencies.

## 2. Empirical Model

### 2.1. Specification

We study the following panel data model of crypto asset log-returns:

$$r_{i,t} = \boldsymbol{\beta}_i' \mathbf{x}_{i,t} + \boldsymbol{\gamma}_i' \mathbf{y}_t + \boldsymbol{\delta}_i' \mathbf{g}_t + \varepsilon_{i,t}, \quad i = 1, \ldots, N; \quad t = 1, \ldots, T. \qquad (1)$$

Here, $r_{i,t}$ denotes the log-return of crypto asset $i$ in week $t$. The vector $\mathbf{x}_{i,t} \in \mathbb{R}^{K_x}$ collects time-varying idiosyncratic risk characteristics, $\mathbf{y}_t \in \mathbb{R}^{K_y}$ represents core market-level risk factors and $\mathbf{g}_t \in \mathbb{R}^{K_g}$ captures macro-financial forces, such as policy uncertainty, geopolitical risk, narrative-driven expectations, and shifts in business and investor sentiment. We refer to these aggregate influences collectively as "economy-wide risk". The slope coefficient vectors $\boldsymbol{\beta}_i$, $\boldsymbol{\gamma}_i$ and $\boldsymbol{\delta}_i$ are allowed to vary freely across crypto assets, accommodating heterogeneous exposures. The term $\varepsilon_{i,t}$ captures purely idiosyncratic noise.

We define $\mathbf{x}_{i,t} \equiv (r_{i,t-1}, \text{ILQ}_{i,t}, \text{VLT}_{i,t})'$, where $r_{i,t-1}$ is the one-period lagged return, and $\text{VLT}_{i,t}$ and $\text{ILQ}_{i,t}$ represent idiosyncratic illiquidity and volatility, respectively for each asset (formal definitions are provided in Section 4), while the corresponding slope coefficients are $\boldsymbol{\beta}_i = (\beta_{1,i}, \beta_{2,i}, \beta_{3,i})'$.[3]

The vector of core market-level financial risk factors is defined as:

$$\mathbf{y}_t \equiv (CMKT_t, \text{CVLT}_t, SMKT_t, \text{SVLT}_t, BMKT_t, \text{BVLT}_t, \\ GMKT_t, \text{GVLT}_t, CSIZE_t, CMOM_t)'. \qquad (2)$$

Here, $CMKT_t$ and $\text{CVLT}_t$ denote the return and volatility of the overall crypto market, respectively, not limited to the digital assets included in our sample. The subsequent components represent returns and volatilities in traditional financial asset classes; namely, stocks (prefixed with "S"), bonds ("B"), and gold ("G").

The use of returns and volatilities in these markets is based on evidence that exposures to these factors are important determinants of crypto asset returns (see Panagiotidis et al. (2019), Anyfantaki et al. (2021), and Hu et al. (2023), among others). We include two additional crypto asset market factors, size ($CSIZE_t$) and momentum ($CMOM_t$), as these have been identified as important return determinants in the literature (Liu et al., 2022). In digital asset markets, the tendency for smaller assets to outperform larger ones – known as the size premium – can be attributed to liquidity differences or to a trade-off where larger, more established crypto assets offer higher "convenience yields" but deliver lower capital gains, as discussed in Cong et al. (2021) and Sockin and Xiong (2023). The momentum premium, where assets that have performed well in the past tend to continue performing well, is also present in crypto markets, consistent with the notion that investors may overreact to information, a pattern similarly documented in traditional equity markets Barberis et al. (1998).

---

[3]To ensure stability of the model, we assume $\sup_{1 \leq i \leq N} |\beta_{1,i}| < 1$.

2.2. Motivation

The specification in Eq. (1) offers a flexible, data-driven framework for analyzing realized return behavior across heterogeneous crypto assets. By incorporating asset-specific frictions, market-level risk, and broader economy-wide forces, the model captures the key sources of both idiosyncratic and systematic variation in returns. It is intentionally descriptive in nature, aiming to explain realized return variation without imposing equilibrium restrictions or assuming a common pricing kernel.

Specifically, the inclusion of $\mathbf{y}_t$, captures market-level risk in the spirit of classical factor pricing models such as the CAPM and APT (Sharpe, 1964; Lintner, 1969; Ross, 1976). In contrast, $\mathbf{g}_t$ reflects broader economy-wide risk factors not already encapsulated by $\mathbf{y}_t$. This aligns with prominent factor-augmented approaches in empirical finance (Connor and Korajczyk, 1986; Bai and Ng, 2002; Ludvigson and Ng, 2007). The asset-specific characteristics $\mathbf{x}_{i,t}$ are motivated by conditional return models that allow expected returns to vary with observable features, such as liquidity and volatility, as emphasized in the literature on cross-sectional return predictability (Ferson and Harvey, 1999; Lettau and Ludvigson, 2001; Liu et al., 2022). Conditional on $\mathbf{y}_t$ and $\mathbf{g}_t$, significant coefficients on $\mathbf{x}_{i,t}$ may signal the presence of short-term risk premia, liquidity-related frictions or limited diversification (e.g., Goetzmann and Kumar (2008)).

The role of $\mathbf{g}_t$ is crucial in this context, reflecting the inherently speculative nature of crypto markets. Fundamentals often exert limited influence on asset valuations, which are instead shaped by broader macro-financial forces such as policy uncertainty, geopolitical risk, and shifts in investor sentiment. As a result, shocks to macro-financial conditions can exert an outsized impact on crypto markets, often even in the absence of asset-specific or market-level news. The component $\mathbf{g}_t$ is designed to capture this diffuse yet pervasive layer of systemic risk, which operates beyond the scope of asset-level fundamentals or idiosyncratic events.

A key challenge stems from the fact that $\mathbf{g}_t$ is inherently latent: macro-financial shocks are typically measured using indicators available only at monthly or quarterly intervals, with almost no data recorded at high frequency.

Complicating matters further, the set of candidate proxies is both large and diverse, spanning measures of macroeconomic conditions, geopolitical tensions, investor sentiment, and media attention. A case in point is "economic uncertainty", which can be proxied by numerous competing indices (Baker et al., 2016; Caldara and Iacoviello, 2022; Ahir et al., 2022; Jurado et al., 2015), often available in multiple variants depending on forecast horizon and geographic coverage. For example, macroeconomic uncertainty may be reported at 1-month, 3-month, and 12-month horizons, while inflation expectations appear at 1-year, 3-year, and 5-year intervals. This raises the empirical question of which metric – and which horizon – is most relevant, especially given the high degree of collinearity among candidate proxies. This phenomenon of factor proliferation – commonly referred to as the "factor zoo" – is well documented in the equity markets literature (Cochrane, 2011; Harvey et al., 2016; Hou et al., 2020). Including too few variables risks omitted variable bias, while including too many introduces multicollinearity and inflates the risk of overfitting.[4] While this problem is already acute in traditional asset pricing, it is arguably even more pronounced in the context of crypto assets, where empirical research is still emerging and the financial ecosystem continues to evolve rapidly. Taken together, these considerations highlight the need for econometric methods that are robust to high dimensionality, latent structure, and the unique dynamics of digital asset markets.

Our divide-and-conquer approach is motivated by the need to address these issues and is summarized graphically in Figure 1.

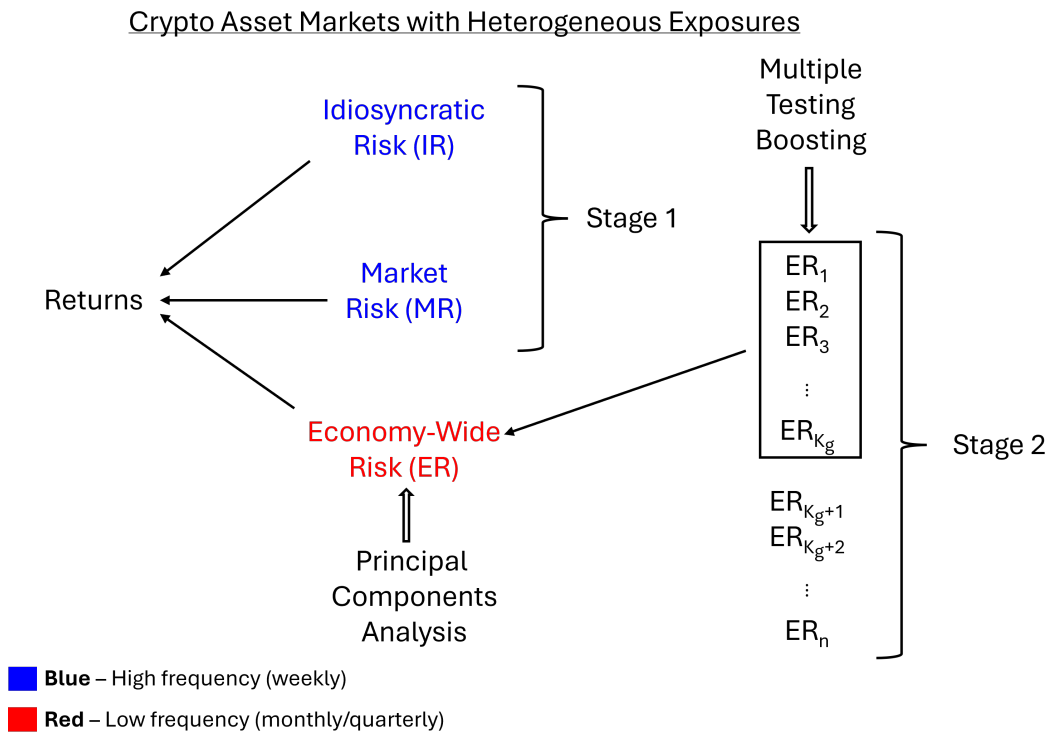

Figure 1: Divide-and-Conquer Approach

In Stage 1, we estimate heterogeneous exposures to high-frequency idiosyncratic and core market-level risk only, with latent economy-wide risk components temporarily absorbed into the model's residuals. In Stage 2, we apply principal components analysis (PCA) to these residuals to semi-parametrically extract the leading principal component. This component represents a linear combination that captures the direction of maximum shared variation across assets within the vector space spanned by these latent common risks. We then apply the multiple testing boosting (MTB) approach of Kapetanios et al. (2025) to identify a sparse subset of low-frequency macro-financial risk proxies, selected from a high-dimensional pool, that best explain the leading principal component.[5] This step effectively translates the one-dimensional leading principal component into a multi-dimensional span of observable risk proxies available at lower frequency. Appendix C presents simulation evidence showing that the combined PCA-MTB method delivers strong finite-sample performance, even when the time dimension is as low as $T = 25$.

---

[4]For example, Harvey et al. (2016) caution against using conventional significance thresholds in asset pricing, arguing that the large number of tested factors and the risk of data mining make many findings statistically spurious.

## 3. Two-Stage Divide-and-Conquer Approach

This section provides the analytical formulation of the divide-and-conquer strategy tailored to model (1). We begin by outlining the key steps in each stage and subsequently elaborate on their implementation and rationale. In what follows, let $\mathbf{M_A} \equiv \mathbf{I}_T - \mathbf{A}(\mathbf{A}'\mathbf{A})^{-1}\mathbf{A}'$ denote the projection matrix onto the orthogonal complement of the column space of any $T \times K_A$ matrix $\mathbf{A}$.

**Stage 1: Observed Risk — Idiosyncratic and Market Components**

— Project out the core market-level risk factors $\mathbf{Y}$ from the semi-endogenous variables of the model $\mathbf{Z}_i$ (to be defined shortly). This yields the projection of $\mathbf{Z}_i$ onto the orthogonal complement of the column space of $\mathbf{Y}$, denoted $\mathbf{M_Y Z}_i$.

— Apply PCA to $\mathbf{M_Y Z}_{i,-\tau}$ for $\tau = 0, 1 \ldots, \zeta$, based on the covariance matrix

$$\frac{1}{NT} \sum_{i=1}^{N} \mathbf{M_Y Z}_{i,-\tau} \mathbf{Z}'_{i,-\tau} \mathbf{M_Y},$$

to extract (up to rotation) the latent common component in $\mathbf{Z}_i$, denoted $\widehat{\mathbf{F}}_{-\tau}$.

— Project $\mathbf{Z}_{i,-\tau}$ onto the orthogonal complement of the column space of $\widehat{\mathbf{F}}_{-\tau}$ using $\mathbf{M}_{\widehat{\mathbf{F}}_{-\tau}} \mathbf{M_Y Z}_{i,-\tau}$. Likewise, eliminate the influence of $\widehat{\mathbf{F}}$ from the outcome equation by computing:
$$\mathbf{M}_{\widehat{\mathbf{F}}} \mathbf{r}_i = \mathbf{M}_{\widehat{\mathbf{F}}} \mathbf{X}_i \boldsymbol{\beta}_i + \mathbf{M}_{\widehat{\mathbf{F}}} \mathbf{Y} \boldsymbol{\gamma}_i + \mathbf{M}_{\widehat{\mathbf{F}}} \mathbf{u}_i.$$

— Construct the instrument matrix
$$\widehat{\mathbf{Z}}_i = \left( \mathbf{M}_{\widehat{\mathbf{F}}} \mathbf{M_Y Z}_i, \ \mathbf{M}_{\widehat{\mathbf{F}}_{-1}} \mathbf{M_Y Z}_{i,-1}, \ \ldots, \ \mathbf{M}_{\widehat{\mathbf{F}}_{-\zeta}} \mathbf{M_Y Z}_{i,-\zeta}, \ \mathbf{Y}, \ \mathbf{Y}_{-1} \right).$$

— Estimate the asset-specific exposures to idiosyncratic and market-level risk, denoted $\boldsymbol{\theta}_i = (\boldsymbol{\beta}'_i, \boldsymbol{\gamma}'_i)'$, using Instrumental Variables (IV).

---

[5]Simulation results in Kapetanios et al. (2025) suggest that MTB performs well in finite samples, outperforming both Lasso and Adaptive Lasso regressions.

**Stage 2: Latent Risk — Economy-Wide Common Components**

— Extract the leading principal component from the IV residuals $\widehat{\mathbf{u}}_i = \mathbf{r}_i - \mathbf{C}_i\widehat{\boldsymbol{\theta}}_i$ using PCA on the residual covariance matrix

$$\frac{1}{NT} \sum_{i=1}^{N} \widehat{\mathbf{u}}_i \widehat{\mathbf{u}}_i'.$$

This leading component, denoted as $\mathbf{e}_1$, represents a linear combination of the model's residuals that captures the direction of maximum shared variation across assets within the vector space spanned by these latent common risks.

— To align with the lower frequency of candidate proxies, aggregate the leading principal component to monthly frequency and denote the resulting series by $\widetilde{\mathbf{e}}_1$. Project $\widetilde{\mathbf{e}}_1$ onto a high-dimensional set of candidate macro-financial risk proxies, denoted $\mathcal{Z}$. Recover the relevant predictors using the multiple testing boosting algorithm of Kapetanios et al. (2025). Let the resulting "design" matrix be $\widetilde{\mathbf{G}}$.

— (Optional) Evaluate the relevant contribution of each selected predictor by regressing $\widetilde{\mathbf{e}}_1$ on $\widetilde{\mathbf{G}}$ and applying the Shapley-Owen decomposition (Shapley, 1953) to the corresponding $R^2$ statistic.

— Estimate each asset's exposures to the identified macro-financial risk proxies by regressing each IV residual vector $\widehat{\mathbf{u}}_i$ on $\widetilde{\mathbf{G}}$. Denote the resulting coefficient vector $\widehat{\boldsymbol{\delta}}_i$.

As a final step, asset-specific exposures to idiosyncratic, market-level, and economy-wide risk, $\widehat{\boldsymbol{\theta}}_i$ and $\widehat{\boldsymbol{\delta}}_i$ respectively, can be aggregated using a Mean Group estimator to uncover structured patterns of heterogeneity across crypto asset categories.

Stage 1 extends the defactored IV framework of Norkute et al. (2021) and Cui et al. (2022) by incorporating observed common factors, $\mathbf{Y}$, into the estimation process, alongside a latent factor component. This extension enables the joint treatment of individual-specific and common covariates – namely, idiosyncratic and market-level risks – a feature not accommodated by their framework. Stage 2 links the latent common structure extracted from residual variations to lower-frequency risk proxies, combining PCA with the multiple testing boosting variable selection method of Kapetanios et al. (2025).

We start by rewriting model (1) to make explicit the structure of the composite error term:

$$r_{i,t} = \boldsymbol{\beta}_i' \mathbf{x}_{i,t} + \boldsymbol{\gamma}_i' \mathbf{y}_t + u_{i,t}; \quad u_{i,t} = \boldsymbol{\delta}_i' \mathbf{g}_t + \varepsilon_{i,t}, \quad i = 1, \ldots, N; \quad t = 1, \ldots, T, \qquad (3)$$

This reformulation highlights that $\mathbf{g}_t$ is inherently unobserved and, at least initially, it is absorbed through a composite error term, reflecting both the lack of high-frequency data and the uncertainty surrounding which among many potential low-frequency proxies are most relevant.

Stacking the $T$ observations for each $i$ in Eq. (3) yields

$$\mathbf{r}_i = \mathbf{X}_i \boldsymbol{\beta}_i + \mathbf{Y} \boldsymbol{\gamma}_i + \mathbf{u}_i; \quad \mathbf{u}_i = \mathbf{G} \boldsymbol{\delta}_i + \boldsymbol{\varepsilon}_i, \qquad (4)$$

where $\mathbf{r}_i = (r_{i,1}, \ldots, r_{i,T})'$, $\mathbf{X}_i = (\mathbf{x}_{i,1}, \cdots, \mathbf{x}_{i,T})'$, a $T \times K_x(= 3)$ matrix, $\mathbf{Y} = (\mathbf{y}_1, \cdots, \mathbf{y}_T)'$, a $T \times K_y(= 10)$ matrix, $\mathbf{u}_i = (u_{i,1}, \ldots, u_{i,T})'$, $\mathbf{G} = (\mathbf{g}_1, \cdots, \mathbf{g}_T)'$ and $\boldsymbol{\varepsilon}_i = (\varepsilon_{i,1}, \cdots, \varepsilon_{i,T})'$.

A first challenge in estimating Eq. (4) stems from the endogeneity of the idiosyncratic risk factors in $\mathbf{X}_i$. We distinguish between semi-endogenous and endogenous variables: an idiosyncratic risk factor is semi-endogenous if it is contemporaneously correlated with the factor component of the error term, $\boldsymbol{\delta}_i' \mathbf{g}_t$; it is fully endogenous if it is correlated with both this component and the idiosyncratic error $\varepsilon_{i,t}$.

By construction, $r_{i,t-1}$ is semi-endogenous, as it is correlated with the latent factor component $\boldsymbol{\delta}_i' \mathbf{g}_t$ unless $\mathbf{g}_t$ is serially independent.[6] The idiosyncratic risk factors, $\text{ILQ}_{i,t}$ and $\text{VLT}_{i,t}$, are also likely subject to semi-endogeneity. Specifically, these variables may be contemporaneously correlated with economy-wide risk sources, $\mathbf{g}_t$.

A further complication arises in the case of $\text{ILQ}_{i,t}$ due to its potential correlation with the idiosyncratic error term in Eq. (3), $\varepsilon_{i,t}$. Specifically, trading behavior may adjust in response to return movements within the weekly observation window considered here. For instance, large positive returns may attract investor inflows and increase trading activity, thereby enhancing market liquidity. Conversely, sharp negative returns may lead to investor withdrawals, lower participation, and heightened illiquidity.[7] This gives rise to potential reverse causality between returns and liquidity conditions.[8]

To address potential reverse causality in $\text{ILQ}_{i,t}$, we use idiosyncratic trading volume, denoted by $\text{VLM}_{i,t}$, as a readily available instrument. While $\text{VLM}_{i,t}$ is likely to be semi-endogenous, due to possible contemporaneous co-movement with $\mathbf{g}_t$, it is unlikely to be correlated with the idiosyncratic error term $\varepsilon_{i,t}$, conditional on $\text{ILQ}_{i,t}$ and $\text{VLT}_{i,t}$.[9]

We collect the semi-endogenous variables of the model into the $K_z(= 2) \times 1$ vector $\mathbf{z}_{i,t} = (\text{VLM}_{i,t}, \text{VLT}_{i,t})'$. We specify

$$\mathbf{z}_{i,t} = \boldsymbol{\Lambda}_i \mathbf{f}_t + \boldsymbol{\Phi}_i \mathbf{y}_t + \mathbf{v}_{i,t}, \tag{5}$$

where $\mathbf{f}_t \in \mathbb{R}^{K_f}$ denotes a vector of latent economy-wide risk factors potentially influencing $\mathbf{z}_{i,t}$ with corresponding loadings given by the $K_z \times K_f$ matrix $\boldsymbol{\Lambda}_i$; $\boldsymbol{\Phi}_i$ captures the heterogeneous sensitivities of $\mathbf{Y}$ on $\mathbf{z}_{i,t}$; and $\mathbf{v}_{i,t} = (v_{1,i,t}, \ldots, v_{K_z,i,t})'$ represents the $K_z \times 1$ vector of idiosyncratic errors. We distinguish between the latent economy-wide risk factors $\mathbf{f}_t$ in Eq. (5) and $\mathbf{g}_t$ in Eq. (3), as the common sources of variation driving idiosyncratic volatility and illiquidity may differ from those influencing return innovations. In practice, the two sets of factors can be identical, correlated, or mutually orthogonal; maintaining this distinction allows for greater modeling flexibility.

---

[6]Standard transformations commonly used in panel data analysis to eliminate fixed effects are not suitable in the present context. For instance, first-differencing transforms the factor component as $\boldsymbol{\delta}_i' \Delta \mathbf{g}_t$, while the within transformation yields $\boldsymbol{\delta}_i' (\mathbf{g}_t - \bar{\mathbf{g}})$. Unless $\mathbf{g}_t$ is serially independent or time-invariant, the transformed regressor thus remains correlated with the factor component, violating the weak exogeneity condition required for consistent estimation. See Sarafidis and Wansbeek (2012, 2021); Juodis and Sarafidis (2022a) for more details.

[7]In contrast, at higher (e.g., intraday) frequencies, liquidity measures are more likely to reflect exogenous microstructure effects, making contemporaneous feedback less of a concern.

[8]In comparison, reverse causality can be reasonably ruled out for $\text{VLT}_{i,t}$, given the well-established direction of dependence. Canonical asset pricing models, such as the CAPM and its intertemporal extension (ICAPM), posit a unidirectional relationship whereby expected returns rise with return volatility, reflecting compensation for systematic risk. This supports interpreting volatility as a forward-looking risk measure, rather than one contemporaneously driven by return realizations.

[9]We provide empirical evidence supporting the exclusion restriction of $\text{VLM}_{i,t}$ in Section 5.3.

To facilitate exposition of our estimation strategy, we rewrite Eqs. (3) and (5) in compact form as

$$\mathbf{r}_i = \mathbf{C}_i \boldsymbol{\theta}_i + \mathbf{u}_i, \tag{6}$$
$$\mathbf{Z}_i = \mathbf{F}\boldsymbol{\Lambda}_i' + \mathbf{Y}\boldsymbol{\Phi}_i' + \mathbf{V}_i; \tag{7}$$

where $\mathbf{C}_i = (\mathbf{X}_i, \mathbf{Y})$ is the $T \times (K_x + K_y)$ regressor matrix, $\boldsymbol{\theta}_i = (\boldsymbol{\beta}_i', \boldsymbol{\gamma}_i')'$ is the corresponding asset-specific coefficient vector, capturing exposures to idiosyncratic and core market-level risk, $\mathbf{Z}_i = (\mathbf{z}_{i,1}, \ldots, \mathbf{z}_{i,T})'$ is $T \times K_z$, $\mathbf{F} = (\mathbf{f}_1, \cdots, \mathbf{f}_T)'$ is $T \times K_f$, and $\mathbf{V}_i = (\mathbf{v}_{i,1}, \ldots, \mathbf{v}_{i,T})'$.

Our two-stage estimation proceeds as follows. We begin by projecting out the market-level risk factors, $\mathbf{Y}$, from the semi-endogenous variables $\mathbf{Z}_i$. This yields the projection of $\mathbf{Z}_i$ onto the orthogonal complement of the column space of $\mathbf{Y}$, denoted $\mathbf{M_Y Z}_i$.

Next, we extract the residual common components in the projected contemporaneous and lagged values $\mathbf{M_Y Z}_{i,-\tau}$, for $\tau = 0, \ldots, \zeta_z$, using PCA. These components are estimated based on the empirical covariance matrix $(NT)^{-1} \sum_{i=1}^{N} \mathbf{M_Y Z}_{i,-\tau} \mathbf{Z}_{i,-\tau}' \mathbf{M_Y}$ (up to rotation), and are denoted by $\widehat{\mathbf{F}}_{-\tau}$.

Subsequently, we project $\mathbf{Z}_{i,-\tau}$ onto the orthogonal complement of the column space of $\widehat{\mathbf{F}}_{-\tau}$ using $\mathbf{M}_{\widehat{\mathbf{F}}_{-\tau}} \mathbf{M_Y Z}_{i,-\tau}$. Likewise, we eliminate the influence of $\widehat{\mathbf{F}}$ from the outcome Equation (6). Specifically, we compute $\mathbf{M}_{\widehat{\mathbf{F}}_{-\tau}} \mathbf{M_Y Z}_{i,-\tau}$ for each lag $\tau$, and transform the outcome equation to remove $\widehat{\mathbf{F}}$ using the following transformation:

$$\mathbf{M}_{\widehat{\mathbf{F}}} \mathbf{r}_i = \mathbf{M}_{\widehat{\mathbf{F}}} \mathbf{X}_i \boldsymbol{\beta}_i + \mathbf{M}_{\widehat{\mathbf{F}}} \mathbf{Y} \boldsymbol{\gamma}_i + \mathbf{M}_{\widehat{\mathbf{F}}} \mathbf{u}_i = \mathbf{M}_{\widehat{\mathbf{F}}} \mathbf{C}_i \boldsymbol{\theta}_i + \mathbf{M}_{\widehat{\mathbf{F}}} \mathbf{u}_i. \tag{8}$$

**Remark 1.** It is important to project $\mathbf{Y}$ out of $\mathbf{Z}_i$ prior to latent factor estimation; otherwise, $\widehat{\mathbf{F}}$ may absorb components of $\mathbf{Y}$, confounding observed and unobserved variation and undermining identification of $\boldsymbol{\gamma}_i$. This critical projection step is not implemented in Norkute et al. (2021) and Cui et al. (2022), as their methodology does not account for the presence of observed factors in the model or instrument set.

The set of instruments is subsequently constructed as

$$\widehat{\mathbf{Z}}_i = \left( \mathbf{M}_{\widehat{\mathbf{F}}} \mathbf{M_Y Z}_i, \ \mathbf{M}_{\widehat{\mathbf{F}}_{-1}} \mathbf{M_Y Z}_{i,-1}, \ \ldots, \ \mathbf{M}_{\widehat{\mathbf{F}}_{-\zeta}} \mathbf{M_Y Z}_{i,-\zeta}, \ \mathbf{Y}, \ \mathbf{Y}_{-1} \right). \tag{9}$$

Projecting $\mathbf{M_Y Z}_{i,-\tau}$ onto the orthogonal complement of the column space of $\widehat{\mathbf{F}}$ eliminates the endogenous component of the semi-endogenous variables, thereby ensuring that the resulting instruments satisfy the exogeneity condition. In addition, the transformation in Eq. (8) eliminates the portion of the latent common component that is shared between $\mathbf{Z}_i$ and $\mathbf{u}_i$ prior to estimation, thereby enhancing efficiency of the asset-specific IV estimator.

Using the instruments in Eq. (9), the asset-specific IV estimator of $\boldsymbol{\theta}_i$ is defined as follows:

$$\widehat{\boldsymbol{\theta}}_i = \left( \widehat{\mathbf{A}}_{i,T}' \widehat{\mathbf{B}}_{i,T}^{-1} \widehat{\mathbf{A}}_{i,T} \right)^{-1} \widehat{\mathbf{A}}_{i,T}' \widehat{\mathbf{B}}_{i,T}^{-1} \widehat{\mathbf{c}}_{i,T}, \tag{10}$$

where

$$\widehat{\mathbf{A}}_{i,T} = \frac{1}{T}\widehat{\mathbf{Z}}'_i \mathbf{M}_{\widehat{\mathbf{F}}} \mathbf{C}_i; \quad \widehat{\mathbf{B}}_{i,T} = \frac{1}{T}\widehat{\mathbf{Z}}'_i \mathbf{M}_{\widehat{\mathbf{F}}} \widehat{\mathbf{Z}}_i; \quad \widehat{\mathbf{c}}_{i,T} = \frac{1}{T}\widehat{\mathbf{Z}}'_i \mathbf{M}_{\widehat{\mathbf{F}}} \mathbf{y}_i. \qquad (11)$$

Under standard regularity conditions and the validity of the instruments, $\widehat{\boldsymbol{\theta}}_i$ achieves $\sqrt{T}$-consistency. This result holds as long as $T$ is large and the number of instruments is held fixed, ensuring that the sample moment conditions converge uniformly to their population counterparts. Intuitively, since the latent common components are projected out and the defactored variables are valid instruments, the estimation problem reduces to a standard IV setting with asymptotically vanishing bias and variance of order $1/T$, yielding consistency at rate $\sqrt{T}$.

To be more specific, it is sufficient to make the following assumptions, where $||\mathbf{A}|| = \sqrt{\operatorname{tr}[\mathbf{A}'\mathbf{A}]}$ and $\operatorname{tr}[\mathbf{A}]$ denote the trace and Frobenius (Euclidean) norm of the matrix $\mathbf{A}$, respectively, and $\Delta$ is a finite positive constant.

**Assumption 1 (idiosyncratic error in y)**: $\varepsilon_{i,t}$ is distributed independently across $i$ and $t$, with $E(\varepsilon_{i,t}) = 0$, $E(\varepsilon_{i,t}^2) = \sigma_{\varepsilon,i,t}^2$, and $E|\varepsilon_{i,t}|^{8+\delta} \leq \Delta < \infty$ for a small positive constant $\delta$.

**Assumption 2 (idiosyncratic error in Z)**: (i) $v_{k,i,t}$ is distributed independently across $i$ and group-wise independent from $\varepsilon_{i,t}$; (ii) $E(v_{k,i,t}) = 0$ and $E|v_{k,i,t}|^{8+\delta} \leq \Delta < \infty$; (iii) $T^{-1} \sum_{s=1}^{T} \sum_{t=1}^{T} E|v_{k,i,s}v_{k,i,t}|^{1+\delta} \leq \Delta < \infty$; (iv) $E\left|N^{-1/2}\sum_{i=1}^{N}[v_{k,i,s}v_{k,i,t} - E(v_{k,i,s}v_{k,i,t})]\right|^4 \leq \Delta < \infty$ for every $k$, $t$ and $s$; (v) $N^{-1}T^{-2}\sum_{i=1}^{N}\sum_{t=1}^{T}\sum_{s=1}^{T}\sum_{r=1}^{T}\sum_{w=1}^{T}|\operatorname{cov}(v_{k,i,s}v_{k,i,t}, v_{k,i,r}v_{k,i,w})| \leq \Delta < \infty$; and (vi) the largest eigenvalue of $E(\mathbf{v}_{k,i}\mathbf{v}'_{k,i})$ is bounded uniformly for every $k$, $i$ and $T$.

**Assumption 3 (factors)**: $\mathbf{f}_t = \boldsymbol{C}_x(L)\mathbf{q}_{f,t}$ and $\mathbf{g}_t = \boldsymbol{C}_y(L)\mathbf{q}_{g,t}$, where $\boldsymbol{C}_x(L)$ and $\boldsymbol{C}_y(L)$ are absolutely summable, $\mathbf{q}_{f,t} \sim i.i.d.(\mathbf{0}, \boldsymbol{\Sigma}_f)$ and $\mathbf{g}_t \sim i.i.d.(\mathbf{0}, \boldsymbol{\Sigma}_g)$, with $\boldsymbol{\Sigma}_f$ and $\boldsymbol{\Sigma}_g$ positive definite matrices. Each element of $\mathbf{q}_{f,t}$ and $\mathbf{q}_{g,t}$ has finite fourth-order moments and all are group-wise independent from $\mathbf{v}_{i,t}$ and $\varepsilon_{i,t}$.

**Assumption 4 (factor loadings)**: $\boldsymbol{\Lambda}_i \sim i.i.d.(\mathbf{0}, \boldsymbol{\Sigma}_\Lambda)$, $\boldsymbol{\Phi}_i \sim i.i.d.(\mathbf{0}, \boldsymbol{\Sigma}_\Phi)$, $\boldsymbol{\delta}_i \sim i.i.d.(\mathbf{0}, \boldsymbol{\Sigma}_\delta)$, where $\boldsymbol{\Sigma}_\Lambda$, $\boldsymbol{\Sigma}_\Phi$ and $\boldsymbol{\Sigma}_\delta$ are positive definite matrices, and each element of $\boldsymbol{\Lambda}_i$, $\boldsymbol{\Phi}_i$ and $\boldsymbol{\delta}_i$ has finite fourth-order moments. $\boldsymbol{\Lambda}_i$, $\boldsymbol{\Phi}_i$ and $\boldsymbol{\delta}_i$ are independent groups from $\varepsilon_{i,t}$, $\mathbf{v}_{i,t}$, $\mathbf{q}_{f,t}$ and $\mathbf{q}_{g,t}$.

**Assumption 5 (random coefficients)**: (i) $\boldsymbol{\theta}_i = \boldsymbol{\theta} + \boldsymbol{\eta}_i$, $\boldsymbol{\eta}_i \sim i.i.d.(\mathbf{0}, \boldsymbol{\Sigma}_\eta)$, where $\boldsymbol{\Sigma}_\eta$ is a fixed positive definite matrix; (ii) $\boldsymbol{\eta}_i$ is independent of $\boldsymbol{\Lambda}_i$, $\boldsymbol{\Phi}_i$, $\boldsymbol{\delta}_i$, $\varepsilon_{i,t}$, $\mathbf{v}_{i,t}$, $\mathbf{q}_{f,t}$ and $\mathbf{q}_{g,t}$; and (iii) $\boldsymbol{\eta}_i$ satisfies the tail bound:

$$P(|\eta_{ir}| > z) \leq 2\exp\left(-\frac{1}{2} \times \frac{z^2}{a + bz}\right)$$

for all $z$ (and all $i$) and fixed $a, b > 0$, where $\eta_{i,r}$ is the $r^{\text{th}}$ element of $\boldsymbol{\eta}_i$ for $2 \leq r \leq k+1$.

**Assumption 6 (moment condition)**: (i) $E\|\boldsymbol{\eta}_i\|^4 \leq \Delta$; (ii) $E\|T^{-1/2}\mathbf{V}'_i\mathbf{F}\|^4 \leq \Delta$; and (iii) $E\|N^{-1/2}T^{-1/2}\sum_{k=1}^{k}\sum_{j=1}^{N}(\mathbf{V}'_i\mathbf{v}_{kj} - E(\mathbf{V}'_i\mathbf{v}_{kj}))\boldsymbol{\delta}'_{k,j}\|^4 \leq \Delta$. In addition, (iv) $E(T^{-1/2}\sum_{k=1}^{k}\sum_{t=1}^{T}(v_{kit}^2 - E(v_{k,i,t}^2)))^2 \leq \Delta$.

**Assumption 7** (identification of $\boldsymbol{\theta}_i$): $\mathbf{A}_i = p\lim_{T \to \infty} \widehat{\mathbf{A}}_{i,T}$ has full column rank, $\mathbf{B}_i = p\lim_{T \to \infty} \widehat{\mathbf{B}}_{i,T}$, and $\boldsymbol{\Sigma}_i = p\lim_{T \to \infty} T^{-1} \mathbf{Z}_i' \mathbf{M_F} \mathbf{u}_i \mathbf{u}_i' \mathbf{M_F} \mathbf{Z}_i'$ are positive definite, uniformly.

Theorem 1 below establishes that, despite the endogeneity of idiosyncratic risk, asset-specific exposures can be consistently estimated, and standard inference remains valid under the aforementioned regularity conditions.

**Theorem 1.** *Consider the model in Eqs. (6)–(7), under Assumptions 1–7. Then, as $(N, T) \xrightarrow{j} \infty$ such that $N/T \to c$ with $0 < c < \infty$, for each $i$,*

$$\sqrt{T}\left(\widehat{\boldsymbol{\theta}}_i - \boldsymbol{\theta}_i\right) \xrightarrow{d} N\left(\mathbf{0}, (\mathbf{A}_i' \mathbf{B}_i^{-1} \mathbf{A}_i)^{-1} \mathbf{A}_i' \mathbf{B}_i^{-1} \boldsymbol{\Sigma}_i \mathbf{B}_i^{-1} \mathbf{A}_i (\mathbf{A}_i' \mathbf{B}_i^{-1} \mathbf{A}_i)^{-1}\right), \quad (12)$$

*where $\widehat{\boldsymbol{\theta}}_i$ is defined in Eq. (10), and $\mathbf{A}_i$, $\mathbf{B}_i$ and $\boldsymbol{\Sigma}_i$ are defined in Assumption 7.*

In Stage 2 of our approach, we shift focus to estimating exposures to latent economy-wide risk that remains after controlling for idiosyncratic and financial market-level factors. Specifically, we begin by obtaining the IV residuals from the asset-specific estimations, defined as

$$\widehat{\mathbf{u}}_i = \mathbf{r}_i - \mathbf{C}_i \widehat{\boldsymbol{\theta}}_i. \quad (13)$$

We then extract the leading principal component, $\mathbf{e}_1$, by applying PCA to the empirical covariance matrix $(NT)^{-1} \sum_{i=1}^{N} \widehat{\mathbf{u}}_i \widehat{\mathbf{u}}_i'$. To align with the lower frequency of the candidate proxies, we aggregate this component to monthly frequency, denoting the resulting series as $\widetilde{\mathbf{e}}_1$.

Subsequently, we apply the multiple testing boosting (MTB) algorithm of Kapetanios et al. (2025) to $\widetilde{\mathbf{e}}_1$, using a high-dimensional set of candidate macro-financial risk proxies denoted by $\boldsymbol{\mathcal{Z}} \in \mathbb{R}^{T \times n}$, where $\boldsymbol{\mathcal{Z}} = (\boldsymbol{\mathcal{Z}}_1, \ldots, \boldsymbol{\mathcal{Z}}_n)$. The algorithm returns a selected index set $S \subset \{1, \ldots, n\}$, and the corresponding design matrix $\widetilde{\mathbf{G}} \in \mathbb{R}^{T \times |S|}$, formed by the columns of $\boldsymbol{\mathcal{Z}}$ indexed by $S$, where $|S|$ denotes the cardinality of the selected set. The methodological details of MTB are provided in Appendix C.2.

To assess the relative importance of each selected predictor, we regress $\widetilde{\mathbf{e}}_1$ on $\widetilde{\mathbf{G}}$ and apply the Shapley-Owen decomposition (Shapley, 1953) to the resulting $R^2$ statistic. This yields a robust, model-agnostic measure of each variable's marginal contribution to explanatory power.

Finally, we regress each asset's IV residuals $\widehat{\mathbf{u}}_i$ on $\widetilde{\mathbf{G}}$ to estimate asset-specific exposures to the identified macro-financial drivers:

$$\widehat{\mathbf{u}}_i = \widetilde{\mathbf{G}} \, \boldsymbol{\delta}_i + \boldsymbol{\xi}_i, \quad (14)$$

where $\boldsymbol{\xi}_i$ is a residual component.

The PCA-MTB procedure thus yields a rich and interpretable decomposition of residual return variation into observable drivers of risk, including uncertainty, policy dynamics and investor sentiment. This enhances both the economic content and the policy relevance of the model, and helps identify which observed factors are most closely associated with the latent risk structure. The following result formalizes the sufficiency of the combined PCA-MTB procedure for recovering latent economy-wide risk components from residual return variation.

**Proposition 1.** *Let $\widetilde{\mathbf{e}}_1$ denote the leading principal component extracted from the residual covariance structure, and let $\boldsymbol{\mathcal{Z}} \in \mathbb{R}^{T \times n}$ be the full set of candidate macro-financial risk proxies, among which a sparse subset governs the true data-generating process for $\widetilde{\mathbf{e}}_1$. Under regularity conditions specified in Appendix B, the multiple testing boosting (MTB) procedure applied to the regression of $\widetilde{\mathbf{e}}_1$ on $\boldsymbol{\mathcal{Z}}$, consistently identifies the true set of relevant risk factors, with probability approaching one as $T \to \infty$.*

*Proof.* See Appendix B. □

Proposition 1 establishes that regressing the leading principal component on the full set of candidate proxies and applying MTB recovers the true set of relevant economy-wide risk factors, with probability approaching one. Appendix C explores the finite-sample properties of PCA-MTB with $T \in \{25, 50, 100\}$, and shows that the method performs robustly across a range of scenarios. In particular, it achieves high selection accuracy and stability with high probability, even in high-dimensional settings with limited time-series observations.

Once $\widehat{\boldsymbol{\theta}}_i$ and $\widehat{\boldsymbol{\delta}}_i$ are obtained, they are aggregated into cross-sectional averages over financially meaningful groups, yielding heterogeneity-robust group-level inference.

In particular, the Mean Group (MG) estimator of the population average $\boldsymbol{\theta}$ is

$$\widehat{\boldsymbol{\theta}} = \frac{1}{N} \sum_{i=1}^{N} \widehat{\boldsymbol{\theta}}_i. \tag{15}$$

The following proposition establishes the limiting distribution of the MG estimator:

**Proposition 2.** *Consider the model in Eqs. (6)–(7), under Assumptions 1–7. Then, as $(N, T) \xrightarrow{j} \infty$ such that $N/T \to c$ with $0 < c < \infty$,*

$$\sqrt{N} \left( \widehat{\boldsymbol{\theta}} - \boldsymbol{\theta} \right) \xrightarrow{d} N \left( \mathbf{0}, \boldsymbol{\Sigma}_\eta \right); \tag{16}$$

*and*

$$\widehat{\boldsymbol{\Sigma}}_\eta - \boldsymbol{\Sigma}_\eta \xrightarrow{p} \mathbf{0}, \tag{17}$$

*where*

$$\widehat{\boldsymbol{\Sigma}}_\eta = \frac{1}{N-1} \sum_{i=1}^{N} \left( \widehat{\boldsymbol{\theta}}_i - \widehat{\boldsymbol{\theta}} \right) \left( \widehat{\boldsymbol{\theta}}_i - \widehat{\boldsymbol{\theta}} \right)'. \tag{18}$$

*Proof.* This follows directly from Norkute et al. (2021). □

Similarly, the MG estimator of the population average $\boldsymbol{\delta}$ is given by $\widehat{\boldsymbol{\delta}} = \frac{1}{N} \sum_{i=1}^{N} \widehat{\boldsymbol{\delta}}_i$. Under a similar set of regulatory conditions, $\widehat{\boldsymbol{\delta}}$ is $\sqrt{N}$-consistent and asymptotically normally distributed, with variance-covariance matrix as in Eq. (18).

**Remark 2.** As an alternative for Stage 2, one can apply asset-specific Lasso regressions directly on the IV residuals $\hat{\mathbf{u}}_i$ using the high-dimensional set $\mathcal{Z}$. While straightforward in principle, this approach has several limitations. First, it is computationally demanding, as it requires tuning and solving $N$ separate penalized regressions. In our Monte Carlo analysis (see Appendix C), conducting 2,000 replications of the model with $N = T = 50$ took approximately 1 minute using PCA-MTB, compared to over 14.5 hours for the individual Lasso ("i-Lasso") approach, due to the need to perform separate model selection and estimation for each asset. Second, the selected predictors may vary substantially across assets, complicating comparisons and group-level interpretation. Third, and more critically for Mean Group estimation, zero coefficients may reflect either true absence of exposure or failure to meet the penalization threshold. Averaging over such a mixed distribution leads to a non-standard Mean Group estimator that is difficult to interpret. Post-Lasso OLS can reduce this issue, but often results in overly dense models, undermining the purpose of regularization.

## 4. Data

This section applies our approach to weekly observations spanning the period from 30 December 2019 to 26 December 2022. The use of weekly frequency is driven by data availability for three key crypto market factors, which are only observed at weekly intervals up to early 2023. The dataset combines information at different sampling frequencies, including daily and weekly price and volume information for a broad sample of digital assets that collectively account for more than 80% of the total crypto market capitalization, sourced from Yahoo Finance. This creates a panel with $N = 40$ assets and $T = 157$ weekly observations. As of January 2025, the selected assets represented over 83% of total crypto market capitalization. The market exhibits high concentration, with Bitcoin, Ethereum, XRP, and Tether collectively representing approximately 74% of the total. Nevertheless, several other assets in the sample play important roles within the broader crypto ecosystem.

Based on the daily and weekly data, we compute the following asset-specific metrics:

— Log returns are computed with successive weekly closing values (Monday prices) as: $r_{i,t} = \ln(p_{i,t}) - \ln(p_{i,t-1})$, where $p_{i,t}$ denotes the price of asset $i$ in week $t$;

— Garman Klass volatility is computed using weekly high, low, opening, and closing prices – denoted $p_{i,t}^{(h)}, p_{i,t}^{(l)}, p_{i,t}^{(o)}, p_{i,t}^{(c)}$, respectively – according to: $VLT_{i,t} = 0.5 \left(\ln(p_{i,t}^{(h)}/p_{i,t}^{(l)})\right)^2 - (2\ln(2) - 1)\left(\ln(p_{i,t}^{(c)}/p_{i,t}^{(o)})\right)^2$;

— Amihud illiquidity is computed using daily data. Specifically, we take the ratio of the absolute daily log return of asset $i$ in day $d$, $|r_{i,d}|$ to its daily volume, $VLM_{i,d}$ and average over the week (7 days): $ILQ_{i,t} = 10^6 \frac{1}{7} \sum_{d=1}^{7} \frac{|r_{i,d}|}{VLM_{i,d}}$

Although our main empirical results rely on the Garman–Klass volatility estimator (Garman and Klass, 1980), alternative measures based on historical price data – such as those proposed by Parkinson (1980), Rogers and Satchell (1991), and Yang and Zhang (2000) –yield broadly similar outcomes.

Liquidity is a key consideration in crypto markets, as it directly affects price stability and market efficiency. For this reason, we employ the Amihud illiquidity ratio (Amihud, 2002), which has been shown to be a reliable proxy for liquidity in crypto assets (Brauneis et al., 2021).

We also compute a measure of overall crypto market volatility ($CVLT$), defined as the capitalization-weighted average of individual crypto-specific Garman-Klass volatilities. As previously mentioned, the crypto market is highly concentrated; therefore, the resulting average crypto market volatility serves as a reliable proxy for the volatility of the entire market.

Next, we incorporate a set of core market-level risk factors specific to digital asset markets, following the framework proposed by Liu et al. (2022). Specifically, we include the crypto market factor ($CMKT$), the crypto market size factor ($CSIZE$), and the crypto market momentum factor ($CMOM$). These market-level factors correspond to the entire cryptocurrency market and are obtained from the Yukun Liu website (https://www.yukunliu.com/research/).

In addition, we draw from three traditional asset classes to construct a broader set of representative financial risk factors. These include: log returns on the NASDAQ stock market ($SMKT$); the CBOE Volatility Index ($SVLT$) as a measure of market expectations for near-term volatility based on S&P 500 option prices; log returns on 10-year U.S. Treasury bonds ($BMKT$); the MOVE Index ($BVLT$), which captures market-implied volatility in the U.S. Treasury bond market; gold log returns ($GMKT$); and the CBOE Gold ETF Volatility Index ($GVLT$), which reflects the expected 30-day implied volatility of returns on the SPDR Gold Shares ETF. These data were retrieved from Thomson Reuters Eikon.

Table 1 presents summary statistics for the idiosyncratic and market-level risk variables. The between standard deviation captures variation across assets, while the within standard deviation reflects time-series variation within each asset. For market-level risk variables, which are constant across assets at each point in time, the between standard deviation is zero.

To examine systematic heterogeneity in risk exposures, we classify crypto assets into several economically meaningful subgroups: green vs. non-green assets, stablecoins vs. non-stablecoins, and DeFi vs. non-DeFi assets.

Green digital assets prioritize energy efficiency and environmental sustainability. Specifically, green crypto assets typically utilize a less energy-intensive consensus mechanism known as "proof of stake" (PoS), "Proof-of-Space-and Time" (PoST), and "Proof-of-Authority" (PoA), whereas non-green crypto assets often rely on the more energy-intensive "proof of work" (PoW) system.[10] Our sample includes fifteen green crypto assets: ADA, ALGO, ATOM, BNB, CRO, EOS, FIL, IOTA, IRIS, KAVA, RUNE, THETA, XLM, XRP, XTZ.

Stablecoins are designed to maintain a stable value relative to a reference asset, typically a fiat currency such as the US dollar or commodities like gold. Their primary aim is to

---

[10]On September 15, 2022, "The Ethereum Merge" took place, marking the transition of Ethereum from PoW to PoS in order to improve the scalability of the Ethereum blockchain. As a result of this change, many crypto assets operating on the Ethereum blockchain became "green" after that date. Our sample covers the period from 2020 to 2022, indicating that 91% of our observations occur before "The Merge". This justifies our classification of ETH and other crypto assets based on the Ethereum blockchain during this period.

Table 1: Summary Statistics for Idiosyncratic and Market-Level Risk Factors

| Variable | Mean | Standard Deviation | | | Min | Max |
|---|---|---|---|---|---|---|
| | | Overall | Within | Between | | |
| r | 0.004 | 0.179 | 0.179 | 0.009 | -4.444 | 1.509 |
| VLT | 0.147 | 0.134 | 0.114 | 0.072 | 0.000 | 3.390 |
| ILQ | 48.458 | 1153.2 | 1112.789 | 306.435 | 0.000 | 57929 |
| CMKT | 0.011 | 0.098 | 0.098 | 0.000 | -0.353 | 0.311 |
| CVLT | 0.065 | 0.039 | 0.039 | 0.000 | 0.0021 | 0.311 |
| CSIZE | 0.018 | 0.066 | 0.066 | 0.000 | -0.147 | 0.305 |
| CMOM | 0.015 | 0.066 | 0.066 | 0.000 | -0.179 | 0.381 |
| SMKT | 0.001 | 0.038 | 0.038 | 0.000 | -0.141 | 0.125 |
| SVLT | 24.899 | 8.948 | 8.948 | 0.000 | 12.100 | 82.690 |
| BMKT | -0.001 | 0.010 | 0.010 | 0.000 | -0.033 | 0.038 |
| BVLT | 78.747 | 32.810 | 32.810 | 0.000 | 36.970 | 156.950 |
| GMKT | 0.001 | 0.022 | 0.022 | 0.000 | -0.090 | 0.077 |
| GVLT | 18.950 | 4.597 | 4.597 | 0.000 | 10.910 | 41.430 |

*Notes:* Standard deviations are reported separately as overall, within, and between. All variables are expressed in levels. ILQ values are divided by $10^6$ for ease of interpretation.

provide price stability, making them more suitable for everyday transactions (as a means of payment) and serving as a reliable store of value. Right now, stablecoins are primarily used for buying or selling crypto assets and facilitating cross-border payments. In our sample, we include six stablecoins: USDT – the fourth largest crypto asset by market capitalization – USDC, BUSD, and DAI, all of which are backed by the US dollar, as well as GLC and PAXG, which are backed by gold.

Finally, our third classification of crypto assets relates to Decentralized Finance (DeFi). DeFi projects are applications, platforms, and protocols built using blockchain technology and smart contracts that offer various financial services in a decentralized manner. These projects aim to provide alternatives to traditional financial instruments and processes. Our sample includes sixteen DeFi-related crypto assets, seven of which are also classified as green crypto assets, and three of which are also stablecoins: ADA, ALGO, ANKR, BNB, CRO, DAI, ETH, FET, FIL, KAVA, MATIC, MKR, PERL, RUNE, USDC, USDT.

Table A1 in Appendix A provides details on each asset, including name, symbol, market capitalization, and classification (stablecoin, green asset, or DeFi-related).

Stage 2 of our divide-and-conquer approach involves projecting dominant principal components onto a high-dimensional set of candidate macro-financial risk factors. To this end, we collect 35 variables that serve as proxies for a range of risk channels, including uncertainty, sentiment, market attention, and global financial conditions. These variables have been carefully assembled from a wide array of influential empirical studies and – so far as we are aware – this is the first study to incorporate such a diverse set of indicators within a single modeling framework. The full set of these variables is listed in Table A2, along with an overview of their definitions and data sources.

This list provided in Table A2 is by no means exhaustive. Additional indices not considered here include the global financial uncertainty index of Caggiano and Castelnuovo (2023), which is only available up to May 2020; the global macroeconomic uncertainty indices proposed by Mumtaz and Theodoridis (2017), Redl (2017), and Carriero et al. (2020), which have not been updated to cover the relevant sample period; and the global financial cycle index of Miranda-Agrippino and Rey (2020), which has only recently been extended to include more recent years.

## 5. Discussion

We estimate asset specific exposures to idiosyncratic and systematic risk, $\boldsymbol{\theta}_i$ and $\boldsymbol{\delta}_i$, using the divide-and-conquer approach described in Section 3. The resulting estimates are then averaged across different asset groups to uncover structured patterns.

### 5.1. Main findings
#### 5.1.1. Full Sample

We start by examining the average estimated exposures to idiosyncratic and market-wide risk factors across the full sample of assets (labeled "Full Sample"). The corresponding results are presented in the first column of Table 2.[11]

The average coefficient on $r_{i,t-1}$ is negative and highly significant, indicating the presence of short-term mean reversion in asset returns. This pattern may reflect underlying market frictions such as limited arbitrage opportunities or profit-taking behavior by investors.

The average coefficient on idiosyncratic volatility ($VLT$) is positive and statistically significant, indicating that higher own-asset volatility tends to be associated with higher realized returns. Although our model is not an asset pricing model per se, this sensitivity could be interpreted as reflecting compensation for bearing idiosyncratic risk that investors are unable or unwilling to diversify away. Similarly, the positive coefficient on $ILQ$ (idiosyncratic illiquidity) points to the presence of an illiquidity premium: less liquid crypto assets are associated with higher returns, consistent with compensation for trading frictions (Zhang et al., 2024).

Turning to core market-level risk factors, we observe strong and significant exposure to the overall crypto market return ($CMKT$). Other systematic factors – including common volatility ($CVLT$), size ($CSIZE$), and momentum ($CMOM$) – also exert statistically significant effects on average, reinforcing the idea that a large portion of return variation is driven by shared, sector-level risk. This finding aligns with previous literature documenting strong return spillovers among crypto assets and highlighting the market's high concentration (Corbet et al., 2018; Aslanidis et al., 2021).

It is worth mentioning that the negative coefficient on $CVLT$ arises because it is estimated conditional on individual volatility ($VLT$); when an asset's own volatility is held constant, an increase in aggregate market volatility makes that asset relatively less risky in a portfolio

---

[11]The results are obtained using the Stata command `xtivdfreg`, developed by Kripfganz and Sarafidis (2021, 2025). For our main results, the matrix of instruments is given by Eq. (9) with $\zeta = 5$. However, our results are robust to alternative choices of $\zeta$, for example, to $\zeta = 1$.

Table 2: Mean Group Exposures to Idiosyncratic and Market-Level Risk across Crypto Asset Groups

|  | (1) Full Sample | (2) Green | (3) non-Green | (4) Stable | (5) non-Stable | (6) DeFi | (7) Non-DeFi |
|---|---|---|---|---|---|---|---|
| $r_{i,t-1}$ | -0.123*** | -0.062*** | -0.160*** | -0.227*** | -0.105*** | -0.142*** | -0.111*** |
|  | (0.023) | (0.025) | (0.032) | (0.087) | (0.022) | (0.047) | (0.024) |
| VLT | 0.383*** | 0.493*** | 0.318** | -0.002 | 0.451*** | 0.450*** | 0.339*** |
|  | (0.090) | (0.070) | (0.137) | (0.019) | (0.102) | (0.105) | (0.134) |
| ILQ | 0.003** | 0.001 | 0.004** | 0.005 | 0.002** | 0.004* | 0.002** |
|  | (0.001) | (0.001) | (0.002) | (0.005) | (0.001) | (0.003) | (0.001) |
| CMKT | 0.641*** | 0.677*** | 0.619*** | 0.109 | 0.734*** | 0.619*** | 0.655*** |
|  | (0.047) | (0.045) | (0.070) | (0.106) | (0.031) | (0.093) | (0.050) |
| CVLT | -1.458*** | -1.743*** | -1.288*** | -0.204 | -1.680*** | -1.616*** | -1.354*** |
|  | (0.175) | (0.096) | (0.270) | (0.182) | (0.178) | (0.247) | (0.242) |
| CSIZE | 0.251*** | 0.370*** | 0.180*** | -0.009 | 0.297*** | 0.332*** | 0.197*** |
|  | (0.045) | (0.033) | (0.065) | (0.017) | (0.048) | (0.070) | (0.056) |
| CMOM | 0.114*** | 0.115*** | 0.113*** | 0.169 | 0.104*** | 0.066** | 0.146*** |
|  | (0.032) | (0.041) | (0.046) | (0.177) | (0.024) | (0.037) | (0.047) |
| SMKT | 0.243*** | 0.318*** | 0.198*** | 0.190 | 0.253*** | 0.304*** | 0.203*** |
|  | (0.048) | (0.081) | (0.059) | (0.171) | (0.049) | (0.080) | (0.059) |
| SVLT | 0.001*** | 0.001* | 0.002*** | 0.000* | 0.002*** | 0.001* | 0.002*** |
|  | (0.000) | (0.001) | (0.000) | (0.000) | (0.000) | (0.001) | (0.001) |
| BMKT | -0.873*** | -1.041*** | -0.772*** | 0.387 | -1.095*** | -1.055*** | -0.751*** |
|  | (0.161) | (0.207) | (0.226) | (0.364) | (0.150) | (0.252) | (0.210) |
| BVLT | -0.001*** | -0.001*** | -0.001*** | -0.000 | -0.001*** | 0.000*** | -0.001*** |
|  | (0.000) | (0.000) | (0.000) | (0.000) | (0.000) | (0.000) | (0.000) |
| GMKT | -0.254*** | -0.366*** | -0.186*** | -0.095 | -0.282*** | -0.148 | -0.324*** |
|  | (0.065) | (0.126) | (0.070) | (0.183) | (0.069) | (0.117) | (0.074) |
| GVLT | 0.000 | 0.000 | 0.000 | 0.000 | 0.000 | -0.002*** | 0.001*** |
|  | (0.000) | (0.001) | (0.001) | (0.000) | (0.001) | (0.001) | (0.001) |
| $N$ | 40 | 15 | 25 | 6 | 34 | 16 | 24 |
| $NT$ | 6040 | 2265 | 3775 | 906 | 5134 | 2416 | 3624 |

*Notes:* Standard errors in parentheses. Significance levels: * 10%, ** 5%, *** 1% (one-tailed). Results for the estimated intercept term are omitted for brevity.

context. As a result, its expected return declines, consistent with the risk–return relationship emphasized in financial theory.

The positive and statistically significant average exposures to traditional equity market returns ($SMKT$) and equity market volatility ($SVLT$) indicate that crypto assets have become increasingly sensitive to broader financial market conditions. These results suggest that crypto assets respond systematically to developments in traditional equity markets—both in terms of return levels and market uncertainty. While earlier studies characterized the digital asset market as relatively insulated from conventional financial systems (Corbet et al., 2018; Aslanidis et al., 2019), our findings point to a marked shift in this relationship. This is consistent with recent evidence suggesting that, since the onset of the COVID-19 pandemic, the relationship between crypto asset returns and traditional equity markets – particularly stock market returns and volatility – has intensified, reflecting the growing institutional presence in the crypto space and its increasing integration with established financial markets (see Iyer, 2022).

Average exposures to bond returns ($BMKT$) and bond market volatility ($BVLT$) are negative and statistically significant, suggesting that when bonds returns increase or bonds become more volatile, investor demand for crypto assets tends to decline—possibly reflecting shifts in risk appetite or flight-to-quality behavior. Crypto returns also exhibit negative average exposures to gold returns ($GMKT$), though the relationship is weaker and less robust. By contrast, gold market volatility ($GVLT$) does not exhibit a significant effect in the full sample.

Overall, these results suggest that both idiosyncratic and market-level risk factors jointly shape the of crypto asset returns.

5.1.2. *Comparative analysis of sub-groups*

To better understand how sensitivities vary across asset types, we turn to a comparative analysis of subgroups: green vs non-green assets, stablecoins vs non-stablecoins, and DeFi vs non-DeFi assets.

Estimates of average exposures within each of these groups are reported in Columns (2)–(7) of Table 2. Statistically significant differences across asset types are highlighted in blue font.

**Green vs non-green crypto assets**

There are important differences in the role of idiosyncratic and systematic risk components between green and non-green crypto assets. The average lagged return coefficient is considerably less negative for green assets (–0.062) than for non-green assets (–0.160), indicating that price corrections occur more slowly in the green segment. This slower adjustment may reflect lower trading activity or reduced price informativeness, possibly due to the niche positioning of green assets or the presence of more passive investor bases.

Green assets are, on average, less exposed to idiosyncratic illiquidity ($ILQ$) than their non-green counterparts, indicating that liquidity constraints are more pronounced among non-green crypto assets. In thinner markets, this distinction may prompt exchanges and investors to approach non-green assets with greater caution due to elevated execution risk and the potential for pricing distortions.

From a market-level perspective, both asset types exhibit substantial and broadly similar sensitivity to overall market returns ($CMKT$). However, average exposures to crypto market volatility ($CVLT$) and market size ($CSIZE$) are notably higher for green assets, suggesting that they are more affected by rising uncertainty in the crypto sector and by size-related dynamics. This may reflect greater information asymmetry, or heightened sensitivity to investor sentiment within the green segment. In practical terms, a risk-averse investor may opt to reduce exposure to green assets during periods of elevated volatility, while maintaining positions in more established non-green assets.

Moreover, green assets exhibit, on average, a stronger negative exposure to gold returns (–0.366) than non-green assets (–0.186). This pattern suggests that green assets may behave more like risk-on assets, underperforming during periods when gold performs well – consistent with flight-to-safety dynamics. In contrast, the weaker gold sensitivity of non-green assets points to a more neutral relationship with safe-haven flows, possibly reflecting greater market recognition or a more diversified investor base.

To assess the relative importance of idiosyncratic versus market-level factors in driving asset returns, we conduct a stylized "controlled experiment" based on Mean Group coefficients and average values from our dataset. Specifically, we evaluate how returns respond to a 10% increase in the mean level of each of four key variables – idiosyncratic volatility ($VLT$), illiquidity ($ILQ$), crypto market volatility ($CVLT$), and crypto size ($CSIZE$) – for green and non-green assets.[12] For green assets, the majority of the total return variation in this scenario (approximately 61%) is attributed to market-level influences, with the remaining 39% driven by idiosyncratic risk. In contrast, for non-green assets, idiosyncratic channels dominate, accounting for over 80% of the total impact, compared to just 19% from market-level influences. These results highlight a clear segmentation in return drivers across categories: while green assets exhibit a higher sensitivity to systematic market dynamics, non-green assets are more sensitive to local, asset-specific frictions. This distinction is informative for risk managers and investors alike, suggesting that green assets may behave more like traditional risk assets, whereas non-green crypto assets may require closer monitoring of liquidity and volatility conditions at the individual asset level.

These findings are consistent with recent theoretical and empirical work on green crypto assets (Pastor et al., 2021, 2020), which argues that green securities may exhibit distinct return-generating processes compared to traditional assets. Specifically, the literature suggests that while green assets have recently delivered strong realized returns, these patterns need not reflect higher expected returns going forward. Instead, the divergence may stem from climate-related risk premiums – if green assets hedge climate risk, they may warrant lower expected returns – and from taste premiums driven by investor preferences for sustainability.

Our empirical results, showing that green assets are relatively more tightly linked to market-level dynamics and less dominated by idiosyncratic frictions, are consistent with this view. These assets appear to behave more like broad risk assets, potentially subject to systematic pricing forces and investor sentiment tied to environmental themes. In contrast, non-green crypto assets remain more exposed to local, idiosyncratic risks, highlighting their distinctive behavior and reduced integration with broader market forces.

---

[12]Let $\widehat{\beta}$ and $\widehat{\gamma}$ denote Mean Group estimates of coefficients, and let $\Delta x$ and $\Delta y$ represent a 10% increase from the group-specific mean of each variable $x$ and $y$. For each group, we compute the absolute contribution of each variable to returns as $|\beta \cdot \Delta x|$ for idiosyncratic variables ($VLT$ and $ILQ$) and $|\gamma \cdot \Delta x|$ for market-level variables ($CVLT$ and $CSIZE$). The sum of all four contributions yields the total effect. The share attributed to idiosyncratic risk is the sum of the two idiosyncratic effects divided by the total.

**Stablecoins vs non-stablecoins**

Starting with the idiosyncratic risk factors, we observe notable heterogeneity across categories. The average coefficient on lagged returns $r_{i,t-1}$ is substantially more negative for stablecoins (–0.227) than for non-stablecoins (–0.105), indicating a stronger tendency toward mean reversion in stablecoins. This is consistent with their design objective of maintaining a pegged value, which may mechanically induce faster correction of deviations from equilibrium. In contrast, non-stablecoins – being more susceptible to speculative dynamics – exhibit weaker autoregressive behavior.

Idiosyncratic volatility ($VLT$) and illiquidity ($ILQ$) also exhibit stronger effects for non-stablecoins, indicating that their returns are more influenced by local frictions and asset-specific microstructure dynamics. For instance, the estimated average coefficient on idiosyncratic volatility ($VLT$) is close to zero and statistically insignificant for stablecoins, but large and highly significant for non-stablecoins (0.451). This indicates that the returns of non-stablecoins are, on average, far more sensitive to asset-specific volatility, reflecting greater exposure to short-term uncertainty or speculative trading. In contrast, stablecoins – designed to maintain price stability – appear largely insulated from such fluctuations. These differences may inform risk management practices, potentially encouraging exchanges or investors to monitor $VLT$ more closely when allocating capital to higher-volatility segments of the crypto market.

On the market side, exposure to the overall crypto market ($CMKT$) is significantly higher for non-stablecoins (0.734) than for stablecoins (0.109). This indicates that non-stablecoins co-move more closely with broad crypto market fluctuations, whereas stablecoins remain relatively insulated. Average sensitivities to other systematic risk factors – such as crypto market volatility ($CVLT$) and traditional bond market variables ($BMKT$, $BVLT$) – also differ markedly between the two groups, with non-stablecoins exhibiting greater responsiveness. This underscores that risk transmission mechanisms vary across asset categories, with non-stablecoins more exposed to broader market fluctuations and shifts in investor sentiment. In particular, the average coefficient on bond market returns ($BMKT$) is strongly negative for non-stablecoins (–1.095), but positive and statistically insignificant for stablecoins (0.387). This contrast suggests that non-stablecoins are more vulnerable to bond market downturns—potentially reflecting shifts in risk appetite or flight-to-quality dynamics—whereas the response of stablecoins to bond returns appears weak or uncertain.

**DeFi vs non-DeFi assets**

The comparison between DeFi and non-DeFi assets reveals several statistically significant differences in exposure to systematic risk factors. First, DeFi assets exhibit greater sensitivity to size ($CSIZE$), with an average coefficient of 0.332, compared to 0.197 for non-DeFi assets. This suggests that returns on DeFi assets are more closely tied to capitalization dynamics – possibly reflecting the outsized influence of dominant protocols or the concentration of liquidity within a small number of major platforms.

Conversely, non-DeFi assets show higher sensitivity to momentum ($CMOM$), with an average coefficient of 0.146 versus 0.066 for DeFi assets. This indicates that non-DeFi returns are, on average, more responsive to shifts in overall market momentum, potentially reflecting greater exposure to sentiment-driven rallies or corrections across the crypto sector.

A substantial difference also emerges in response to gold market returns ($GMKT$). Non-DeFi assets exhibit a much stronger negative exposure (–0.324) compared to DeFi assets (–0.148), suggesting that non-DeFi assets react more sharply to movements in safe-haven assets. This pattern may indicate that non-DeFi assets are more susceptible to episodes of macro-level risk aversion, whereas DeFi assets appear somewhat more insulated – possibly due to their functional integration within decentralized financial protocols.

The two groups diverge in their average exposures to gold volatility (GVLT). DeFi assets display a negative response (–0.002), while non-DeFi assets exhibit a positive sensitivity (0.001). This asymmetry may reflect differences in investor composition and motivation. For instance, heightened gold volatility may be interpreted by DeFi participants as a signal of broader financial instability, prompting risk-off behavior. In contrast, non-DeFi investors may perceive increased gold volatility as part of a safe-haven rotation, thereby raising their relative demand for certain crypto assets.

### 5.2. Analysis of latent, economy-wide factors

We begin by extracting the latent common component in $\widehat{\mathbf{u}}_i$, following the procedure outlined in Section 3. The leading principal component accounts for approximately 43% of the total variance in the model's residual variation, indicating that a large portion of the unexplained variation is driven by this component, consistent with the presence of a pervasive latent risk factor. Since macro-financial uncertainty proxies are available at most at monthly frequency, we aggregate the weekly series in $\mathbf{e}_1$ by averaging them within each calendar month.[13] The resulting aggregated series is denoted by $\widetilde{\mathbf{e}}_1$. This temporal aggregation reduces the time dimension of $\widetilde{\mathbf{e}}_1$ to 36 monthly observations, while the set of uncertainty proxies comprises 35 variables. Together, these dimensions highlight the high-dimensional nature of the estimation problem in Stage 2, where the number of candidate proxies is comparable to the number of time-series observations. This makes it infeasible to include all available proxies simultaneously using conventional estimation techniques, necessitating the use of variable selection or regularization methods.

The multiple testing boosting (MBT) method of Kapetanios et al. (2025) selects nine risk predictors; accordingly, the resulting design matrix $\widetilde{\mathbf{G}}$ has dimension $T \times \widehat{K}_g$, where $\widehat{K}_g = 9$. Specifically, the selected variables are:

— the investor sentiment index ($Sent$) by Baker and Wurgler (2006);

— the Google Trends-based crypto market attention index ($GTC$) by Aslanidis et al. (2022);

— the climate policy uncertainty index ($ClimPolicyU$) by Gavriilidis (2021);

— the global economic policy uncertainty index ($GEPU$) by Baker et al. (2016);

— the world uncertainty index ($WUI$) by Ahir et al. (2022);

---

[13] Using median values delivers similar results.

— the consumer sentiment index (*ConsSent*) by University of Michigan (2025a);

— the policy-related news and stock market volatility index (*PN*) by Baker et al. (2019);

— the uncertainty related to crypto prices (*UCRY price*) by Lucey et al. (2022);

— the uncertainty related to crypto policy (*UCRY policy*) by Lucey et al. (2022).

To assess the relative importance of each predictor, we regress the leading principal component $\widetilde{\mathbf{e}}_1$ on $\widetilde{\mathbf{G}}$ and apply the Shapley-Owen decomposition on the $R^2$ statistic (Shapley, 1953). All nine predictors are statistically significant at the 5% level. The fitted values correlate approximately 81% with $\widetilde{\mathbf{e}}_1$, and the regression yields an $R^2$ of 69%, indicating a high degree of explanatory power. Among the nine variables, the largest individual contributions to the explained variance come from the world uncertainty index (*WUI*, 16.5%), uncertainty related to crypto policy (*UCRY policy*, 15.5%), investor sentiment (*Sent*, 14.5%), and consumer sentiment (*ConsSent*, 13%). The remaining five indices jointly account for the remaining 31%. These results suggest that global and crypto-specific policy uncertainty, along with investor and consumer sentiment, are key drivers of the latent factor underlying return variation.

Finally, we estimate the individual exposures of each asset to the nine selected economy-wide risk indices by estimating Eq. (14). To summarize this rich set of information, we compute average exposures using the Mean Group estimator – both across the full sample and within each of the six predefined subgroups. The results are presented in Table 3. Statistically significant differences in exposures across subgroups are highlighted in blue font, allowing for a clear visual identification of cross-group heterogeneity.

For the full sample, investor and consumer sentiment indices exhibit positive average effects on crypto returns. In contrast, indices capturing world uncertainty, crypto market price uncertainty, and climate policy uncertainty are associated with negative return effects. The Google Trends-based attention index also shows a positive average effect, suggesting that heightened public interest tends to coincide with higher crypto returns.

The comparative analysis across subgroups reveals several consistent patterns. First, green assets and non-stablecoins tend to exhibit larger absolute sensitivities to economy-wide indices than their respective counterparts—non-green assets and stablecoins – particularly in cases where the differences are statistically significant (as highlighted in blue). Second, green assets exhibit notably higher exposures to climate policy uncertainty, with the average coefficient approximately 60% larger in magnitude than those of non-green assets, consistent with their closer alignment to climate-related narratives. Third, DeFi assets display heightened sensitivity to global uncertainty relative to non-DeFi assets, while their remaining exposures are broadly similar across the two groups. Finally, stablecoins appear substantially less responsive to uncertainty, sentiment, and attention-based indices than non-stablecoins, reinforcing the interpretation that they constitute a distinct asset class with reduced exposure to macro-financial risk factors.

### 5.3. Robustness

This subsection presents a series of robustness checks aimed at evaluating the validity of our methodology and the stability of our core results. The first two checks pertain to the

Table 3: Mean Group Exposures to Economy-Wide Factors Across Crypto Asset Groups

|  | (1) Full Sample | (2) Green | (3) non-Green | (4) Stable | (5) non-Stable | (6) DeFi | (7) Non-DeFi |
|---|---|---|---|---|---|---|---|
| Sent | 0.032*** | 0.041*** | 0.027*** | 0.010 | 0.036*** | 0.030*** | 0.034*** |
|  | (0.003) | (0.003) | (0.004) | (0.010) | (0.003) | (0.005) | (0.004) |
| WUI | -0.022*** | -0.024*** | -0.020*** | -0.014 | -0.023*** | -0.029*** | -0.017*** |
|  | (0.003) | (0.005) | (0.005) | (0.014) | (0.003) | (0.005) | (0.005) |
| ConsSent | 0.053*** | 0.053*** | 0.053*** | 0.022 | 0.059*** | 0.050*** | 0.055*** |
|  | (0.005) | (0.004) | (0.007) | (0.022) | (0.004) | (0.008) | (0.006) |
| PN | 0.015*** | 0.016*** | 0.014*** | 0.009 | 0.016*** | 0.013*** | 0.016*** |
|  | (0.002) | (0.003) | (0.003) | (0.009) | (0.002) | (0.003) | (0.003) |
| UCRY Price | -2.692*** | -2.493*** | -2.811*** | -1.264 | -2.944*** | -2.630*** | -2.733*** |
|  | (0.354) | (0.393) | (0.520) | (1.330) | (0.338) | (0.631) | (0.426) |
| UCRY Policy | 2.982 | 2.860 | 3.055 | 0.942 | 3.342*** | 3.079*** | 2.917*** |
|  | (0.343) | (0.427) | (0.492) | (1.014) | (0.332) | (0.684) | (0.359) |
| ClimPolicyU | -0.018*** | -0.024*** | -0.015*** | 0.005 | -0.022*** | -0.019*** | -0.017*** |
|  | (0.002) | (0.003) | (0.003) | (0.005) | (0.002) | (0.004) | (0.003) |
| GEPU | 0.038*** | 0.046*** | 0.033*** | 0.018 | 0.042*** | 0.043*** | 0.035*** |
|  | (0.004) | (0.005) | (0.006) | (0.018) | (0.004) | (0.008) | (0.005) |
| GTC | 0.020*** | 0.018*** | 0.022*** | 0.017 | 0.021*** | 0.023*** | 0.019*** |
|  | (0.003) | (0.003) | (0.004) | (0.016) | (0.002) | (0.004) | (0.005) |

*Notes:* Standard errors in parentheses. Significance levels: * 10%, ** 5%, *** 1% (one-tailed). Statistically significant differences in coefficients between groups are highlighted in blue font. The results are based on monthly regressions of estimated residuals, averaged within each month to match the frequency of the macro-financial proxies. Relevant predictors are identified using the multiple testing boosting algorithm of Kapetanios et al. (2025), applied to the full set of 35 available indices. Results for the estimated intercept term are omitted for brevity.

definition of the estimation sample, while the remaining ones involve modifications to the econometric specification.

We begin by restricting the sample to the period prior to the end of quantitative easing (QE) in monetary policy, re-estimating the model using data from 30/12/2019 to 21/03/2022. In addition, the reduced sample excludes the period of the 2022 crypto market turmoil, characterised by the Terra/LUNA collapse in May 2022, the crypto lending crisis in June–July 2022, and the FTX collapse in November 2022. This reduces the time dimension $T$ to 118 observations, while still providing a sufficient number of time series observations for estimation. This exercise allows us to assess the sensitivity of our benchmark results to the end of QE as well as to the inclusion of these high-volatility episodes. The results are reported in the second column of Table 4 (labeled "During QE"). For comparison, the first column reproduces the full-sample estimates from Column (1) of Table 2. The findings for the QE period are broadly consistent with those based on the full sample. Notable exceptions include the coefficients associated with bond market returns and the crypto market momentum factor. Specifically, during the QE period: (i) realized returns display heightened sensitivity to traditional safe-haven assets, such as government bonds – likely reflecting investors' search for yield in a low-interest-rate environment; and (ii) the influence of the momentum factor diminishes substantially, indicating a reduced role for return persistence in driving asset returns during this policy regime.

We then re-estimate the DeFi subsample results after reclassifying three assets – DAI, USDC, and USDT – from the DeFi to the non-DeFi category. These stablecoins, while commonly used in DeFi transactions, are not explicitly DeFi-native. A comparison of Columns (3) and (4) in Table 4 versus Columns (6) and (7) in Table 2 reveals only minor differences, indicating that the core findings remain robust to this reclassification.

Next, we implement two robustness checks based on alternative econometric specifications. First, we reduce the lag structure of the instrument set defined in Eq. (9), using only contemporaneous and one-period-lagged values, rather than the five-lag baseline. Specifically, we define:

$$\widehat{\mathbf{Z}}_i = \left(\mathbf{M}_{\widehat{\mathbf{F}}_z}\mathbf{M}_{\mathbf{Y}}\mathbf{Z}_i,\ \mathbf{M}_{\widehat{\mathbf{F}}_{z,-1}}\mathbf{M}_{\mathbf{Y}}\mathbf{Z}_{i,-1},\ \mathbf{Y},\ \mathbf{Y}_{,-1}\right). \tag{19}$$

This adjustment increases the number of observations from 6,040 to 6,200. The results, presented in Column (5) (labeled "One Lag"), are broadly in line with the baseline estimates in Column (1), with the only notable difference being a moderate change in the AR(1) coefficient. This suggests that our findings are robust to the choice of lag length in the instrument set.

Finally, we assess the exogeneity of our proposed instrument – trading volume ($VLM$) – by including idiosyncratic trading volume directly in the outcome equation. For $VLM$ to serve as a valid instrument, it should have no direct effect on returns, conditional on $VLT$, $ILQ$, and market-level risk factors. The corresponding estimates, reported in Column (6) (labeled "with VLM"), indicate that trading volume has no statistically significant effect on returns. This finding supports the exclusion restriction and confirms that the instrument affects returns only indirectly, through its impact on idiosyncratic illiquidity.

Table 4: Robustness of Mean Group Exposures to Idiosyncratic and Market-Level Risk

|  | (1) Full Sample | (2) During QE | (3) DeFi V2 | (4) Non-DeFi V2 | (5) One Lag | (6) with VLM |
|---|---|---|---|---|---|---|
| $r_{i,t-1}$ | -0.123*** | -0.138*** | -0.131*** | -0.120*** | -0.075** | -0.149*** |
|  | (0.023) | (0.020) | (0.045) | (0.027) | (0.039) | (0.019) |
| VLT | 0.383*** | 0.496*** | 0.561*** | 0.298*** | 0.388*** | 0.206*** |
|  | (0.090) | (0.078) | (0.107) | (0.121) | (0.090) | (0.085) |
| ILQ | 0.003** | 0.004** | 0.003 | 0.003** | 0.003** | 0.002** |
|  | (0.001) | (0.002) | (0.003) | (0.001) | (0.001) | (0.001) |
| CMKT | 0.641*** | 0.637*** | 0.761*** | 0.582*** | 0.639*** | 0.615*** |
|  | (0.047) | (0.046) | (0.064) | (0.060) | (0.049) | (0.048) |
| CVLT | -1.458*** | -1.648*** | -1.990*** | -1.203*** | -1.451*** | -1.481*** |
|  | (0.175) | (0.153) | (0.179) | (0.230) | (0.175) | (0.163) |
| CSIZE | 0.251*** | 0.256*** | 0.407*** | 0.176*** | 0.244*** | 0.159*** |
|  | (0.045) | (0.041) | (0.071) | (0.051) | (0.042) | (0.043) |
| CMOM | 0.114*** | 0.059* | 0.081** | 0.129*** | 0.117*** | 0.135*** |
|  | (0.032) | (0.039) | (0.045) | (0.043) | (0.032) | (0.034) |
| SMKT | 0.243*** | 0.288*** | 0.370*** | 0.182*** | 0.259*** | 0.274*** |
|  | (0.048) | (0.054) | (0.089) | (0.054) | (0.062) | (0.072) |
| SVLT | 0.001*** | 0.002*** | 0.001* | 0.002*** | 0.002*** | 0.003*** |
|  | (0.000) | (0.001) | (0.001) | (0.001) | (0.001) | (0.000) |
| BMKT | -0.873*** | -2.189*** | -1.302*** | -0.666*** | -0.810*** | -0.630*** |
|  | (0.161) | (0.254) | (0.265) | (0.192) | (0.168) | (0.154) |
| BVLT | -0.001*** | -0.001*** | -0.001*** | -0.001*** | -0.001*** | -0.001*** |
|  | (0.000) | (0.000) | (0.000) | (0.000) | (0.000) | (0.000) |
| GMKT | -0.254*** | -0.169** | -0.193* | -0.283*** | -0.261*** | -0.206*** |
|  | (0.065) | (0.074) | (0.142) | (0.069) | (0.069) | (0.064) |
| GVLT | -0.000 | -0.002*** | -0.002*** | -0.001*** | -0.001 | -0.001 |
|  | (0.000) | (0.001) | (0.001) | (0.000) | (0.001) | (0.001) |
| VLM |  |  |  |  |  | 0.000 |
|  |  |  |  |  |  | (0.000) |
| $N$ | 40 | 40 | 13 | 27 | 40 | 40 |
| $NT$ | 6040 | 4440 | 6040 | 6040 | 6200 | 6040 |

*Notes*: Standard errors in parentheses. Significance levels: * 10%, ** 5%, *** 1% (one-tailed). Results for the estimated intercept term are omitted for brevity.

## 6. Concluding Remarks

The rise of blockchain technology and crypto markets marks a foundational shift in financial infrastructure, accelerating the emergence of decentralized and digitally native asset classes. As these markets grow in scale, complexity, and integration with traditional finance, rigorous empirical research is essential to inform how investors, institutions, and policymakers engage with this evolving landscape.

This paper introduces a novel two-stage divide-and-conquer approach to modeling return behavior in crypto assets. Our study contributes at a conceptual level by offering a coherent framework to disentangle heterogeneous risk exposures across idiosyncratic, market-level, and latent economy-wide components. This decomposition is particularly valuable when analyzing asset classes like crypto, where observable fundamentals can be weak or ill-defined, and traditional asset pricing models might struggle to explain return behavior.

The results of our study carry several important implications. For investors, the observed heterogeneity in return sensitivities across asset categories offers valuable guidance for portfolio construction and risk diversification strategies. From a policy perspective, the increasing responsiveness of crypto returns to movements in traditional financial markets underscores the potential for macroeconomic and regulatory actions to trigger cross-market spillovers. In addition, the behavior of green crypto assets suggests that alignment with climate-related investment themes may heighten vulnerability to shifts in policy or regulatory focus.

Looking ahead, a natural extension of this study is to explore the extent to which digital assets not only respond to macro-financial shocks, but also serve as amplifiers or transmission channels in times of stress. While our analysis focuses on directional exposure — that is, how economy-wide risk propagates to crypto returns — the rapid expansion of digital asset markets, coupled with growing linkages to traditional financial institutions through derivatives, custody services, and tokenized assets, raises questions about potential feedback effects that may amplify systemic vulnerabilities. For instance, heightened volatility in crypto markets could spill over into broader financial conditions via investor deleveraging, correlated asset selloffs, or liquidity disruptions in interconnected platforms. Investigating these dynamics will help policymakers identify new sources of systemic risk and design safeguards that reflect the evolving structure of digital financial markets.

## References

Ahir, H., Bloom, N., and Furceri, D. (2022). The World Uncertainty Index. NBER Working Papers 29763, National Bureau of Economic Research, Inc.

Ahn, S. C. and Horenstein, A. R. (2013). Eigenvalue ratio test for the number of factors. *Econometrica*, 81(3):1203–1227.

Amihud, Y. (2002). Illiquidity and stock returns: cross-section and time-series effects. *Journal of Financial Markets*, 5(1):31–56.

Anyfantaki, S., Arvanitis, S., and Topaloglou, N. (2021). Diversification benefits in the cryptocurrency market under mild explosivity. *European Journal of Operational Research*, 295:378–393.

Aslanidis, N., Bariviera, A. F., and Martínez-Ibañez, O. (2019). An analysis of cryptocurrencies conditional cross correlations. *Finance Research Letters*, 31:130–137.

Aslanidis, N., Bariviera, A. F., and Perez-Laborda, A. (2021). Are cryptocurrencies becoming more interconnected? *Economics Letters*, 199:109725.

Aslanidis, N., Bariviera, A. F., and Óscar G. López (2022). The link between cryptocurrencies and google trends attention. *Finance Research Letters*, 47:102654.

Augustin, P., Rubtsov, A., and Shin, D. (2023). The impact of derivatives on spot markets: Evidence from the introduction of bitcoin futures contracts. *Management Science*, 69:6752–6776.

Bai, J. (2003). Inferential Theory for Factor Models of Large Dimensions. *Econometrica*, 71(1):135–171.

Bai, J. (2009). Panel Data Models with Interactive Fixed Effects. *Econometrica*, 77(4):1229–1279.

Bai, J. and Ng, S. (2002). Determining the number of factors in approximate factor models. *Econometrica*, 70(1):191–221.

Bai, J. and Ng, S. (2006). Confidence intervals for diffusion index forecasts and inference for factor-augmented regressions. *Econometrica*, 74(4):1133–1150.

Baker, M. and Wurgler, J. (2006). Investor sentiment and the cross-section of stock returns. *The Journal of Finance*, 61:1645–1680.

Baker, S. R., Bloom, N., and Davis, S. J. (2016). Measuring Economic Policy Uncertainty. *The Quarterly Journal of Economics*, 131(4):1593–1636.

Baker, S. R., Bloom, N., Davis, S. J., and Kost, K. J. (2019). Policy news and stock market volatility. Working Paper 25720, National Bureau of Economic Research.

Barberis, N., Shleifer, A., and Vishny, R. (1998). A model of investor sentiment. *Journal of Financial Economics*, 49:307–343.

Bariviera, A. F. (2017). The inefficiency of bitcoin revisited: A dynamic approach. *Economics Letters*, 161:1–4.

Board of Governors of the Federal Reserve System (2025). Nominal broad u.s. dollar index [dtwexbgs]. retrieved 02/02/2025.

Brauneis, A., Mestel, R., Riordan, R., and Theissen, E. (2021). How to measure the liquidity of cryptocurrency markets? *Journal of Banking & Finance*, 124:106041.

Bua, G., Kapp, D., Ramella, F., and Rognone, L. (2024). Transition versus physical climate risk pricing in european financial markets: a text-based approach. *The European Journal of Finance*, 30:2076–2110.

Caggiano, G. and Castelnuovo, E. (2023). Global financial uncertainty. *Journal of Applied Econometrics*, 38:432–449.

Caldara, D. and Iacoviello, M. (2022). Measuring Geopolitical Risk. *American Economic Review*, 112(4):1194–1225.

Carriero, A., Clark, T. E., and Marcellino, M. (2020). Assessing international commonality in macroeconomic uncertainty and its effects. *Journal of Applied Econometrics*, 35:273–293.

Cochrane, J. H. (2011). Presidential address: Discount rates. *Journal of Finance*, 66(4):1047–1108.

Coinmarket (2025). Crypto-Currency Market Capitalizations. https://coinmarketcap.com/. Accessed: 2025-01-07.

Cong, L., Karolyi, G. A., Tang, K., and Zhao, W. (2022). Value premium, network adoption, and factor pricing of crypto assets. *SSRN Electronic Journal*.

Cong, L. W., Landsman, W., Maydew, E., and Rabetti, D. (2023a). Tax-loss harvesting with cryptocurrencies. *Journal of Accounting and Economics*, 76:101607.

Cong, L. W., Li, X., Tang, K., and Yang, Y. (2023b). Crypto wash trading. *Management Science*, 69:6427–6454.

Cong, L. W., Li, Y., and Wang, N. (2021). Tokenomics: Dynamic adoption and valuation. *The Review of Financial Studies*, 34:1105–1155.

Cong, W., Harvey, C., Rabetti, D., and Wu, Z.-Y. (2025). An anatomy of crypto-enabled cybercrimes. *Management Science*, 71:3622–3633.

Connor, G. and Korajczyk, R. A. (1986). Performance measurement with the arbitrage pricing theory: A new framework for analysis. *Journal of Financial Economics*, 15(3):373–394.

Corbet, S., Meegan, A., Larkin, C., Lucey, B., and Yarovaya, L. (2018). Exploring the dynamic relationships between cryptocurrencies and other financial assets. *Economics Letters*, 165:28–34.

Cui, G., Norkutė, M., Sarafidis, V., and Yamagata, T. (2022). Two-stage instrumental variable estimation of linear panel data models with interactive effects. *The Econometrics Journal*, 25(2):340–361.

Federal Reserve Bank of Chicago (2025). Chicago fed national financial conditions leverage subindex. retrieved 02/02/2025.

Federal Reserve Bank of Cleveland (2025). 5-year expected inflation [expinf5yr]. retrieved 02/02/2025.

Federal Reserve Bank of St. Louis (2025). St. Louis Fed Financial Stress Index [STLFSI4]. https://fred.stlouisfed.org/series/STLFSI4. Accessed: 2025-01-21.

Ferrari Minesso, M., Mehl, A., and Stracca, L. (2022). Central bank digital currency in an open economy. *Journal of Monetary Economics*, 127:54–68.

Ferson, W. E. and Harvey, C. R. (1999). Conditioning variables and the cross section of stock returns. *The Journal of Finance*, 54(4):1325–1360.

Garman, M. B. and Klass, M. J. (1980). On the Estimation of Security Price Volatilities from Historical Data. *The Journal of Business*, 53(1):67–78.

Gavriilidis, K. (2021). Measuring climate policy uncertainty. *SSRN Electronic Journal*.

Goetzmann, W. N. and Kumar, A. (2008). Equity Portfolio Diversification. *Review of Finance*, 12(3):433–463.

Gordon, S., Li, Z., and Marthinsen, J. (2023). A deep analysis of the economics and finance research on cryptocurrencies. *Economics Letters*, 228:111136.

Hafner, C. M. (2020). Testing for Bubbles in Cryptocurrencies with Time-Varying Volatility. *Journal of Financial Econometrics*, 18(2):233–249.

Harvey, C. R., Liu, Y., and Zhu, H. (2016). . . . and the cross-section of expected returns. *The Review of Financial Studies*, 29(1):5–68.

Hou, K., Xue, C., and Zhang, L. (2020). Replicating anomalies. *The Review of Financial Studies*, 33(5):2019–2133.

Hu, Y., Shen, D., and Urquhart, A. (2023). Attention allocation and cryptocurrency return co-movement: Evidence from the stock market. *International Review of Economics & Finance*, 88:1173–1185.

Hur, K. (2024). Crypto has an ally in the white house. what's ahead for digital currencies in 2025.

Iyer, T. (2022). Cryptic connections: Spillovers between crypto and equity markets. *Global Financial Stability Notes*, 2022:1.

Juodis, A. and Sarafidis, V. (2018). xed t dynamic panel data estimators with multi-factor errors. *Econometric Reviews*, 38(8):893–929.

Juodis, A. and Sarafidis, V. (2022a). An incidental parameters free inference approach for panels with common shocks. *Journal of Econometrics*, 229:19–54.

Juodis, A. and Sarafidis, V. (2022b). A linear estimator for factor-augmented fixed-t panels with endogenous regressors. *Journal of Business and Economic Statistics*, 40(1):1–15.

Jurado, K., Ludvigson, S. C., and Ng, S. (2015). Measuring uncertainty. *American Economic Review*, 105(3):1177–1216.

Kapetanios, G., Sarafidis, V., and Ventouri, A. (2025). Model selection in large dimensional linear regression using sequential multiple testing. Working paper.

Kripfganz, S. and Sarafidis, V. (2021). Instrumental variable estimation of large-t panel data models with common factors. *Stata Journal*, 21(3):1–28.

Kripfganz, S. and Sarafidis, V. (2025). Estimating Spatial Dynamic Panel Data Models with Unobserved Common Factors in Stata. *Journal of Statistical Software*, (forthcoming).

Lambrecht, M., Sofianos, A., and Xu, Y. (2025). Does mining fuel bubbles? an experimental study on cryptocurrency markets. *Management Science*, 71:1865–1888.

Lettau, M. and Ludvigson, S. (2001). Resurrecting the (c)capm: A cross-sectional test when risk premia are time-varying. *Journal of Political Economy*, 109(6):1238–1287.

Lintner, J. (1969). The valuation of risk assets and the selection of risky investments in stock portfolios and capital budgets. *Review of Economics and Statistics*, 51(2):222–224.

Liu, Y. and Tsyvinski, A. (2021). Risks and returns of cryptocurrency. *The Review of Financial Studies*, 34:2689–2727.

Liu, Y., Tsyviski, A., and Wu, X. I. (2022). Common risk factors in cryptocurrency. *The Journal of Finance*, 77:1133–1177. https://doi.org/10.1111/jofi.13119.

Lucey, B. M., Vigne, S. A., Yarovaya, L., and Wang, Y. (2022). The cryptocurrency uncertainty index. *Finance Research Letters*, 45:102147.

Ludvigson, S. C. and Ng, S. (2007). The empirical risk–return relation: A factor analysis approach. *Journal of Financial Economics*, 83(1):171–222.

Miranda-Agrippino, S. and Rey, H. (2020). U.S. monetary policy and the global financial cycle. *The Review of Economic Studies*, 87:2754–2776.

Mumtaz, H. and Theodoridis, K. (2017). Common and country specific economic uncertainty. *Journal of International Economics*, 105:205–216.

Norkute, M., Sarafidis, V., Yamagata, T., and Cui, G. (2021). Instrumental variable estimation of dynamic linear panel data models with defactored regressors and a multifactor error structure. *Journal of Econometrics*, 220(2):416–446.

Panagiotidis, T., Stengos, T., and Vravosinos, O. (2019). The effects of markets, uncertainty and search intensity on bitcoin returns. *International Review of Financial Analysis*, 63:220–242.

Parkinson, M. (1980). The Extreme Value Method for Estimating the Variance of the Rate of Return. *The Journal of Business*, 53(1):61–65.

Pastor, L., Stambaugh, R. F., and Taylor, L. A. (2020). Fund tradeoffs. *Journal of Financial Economics*, 138:614–634.

Pastor, L., Stambaugh, R. F., and Taylor, L. A. (2021). Sustainable investing in equilibrium. *Journal of Financial Economics*, 142:550–571.

Redl, C. (2017). The impact of uncertainty shocks in the united kingdom — bank of england.

Rogers, L. C. G. and Satchell, S. E. (1991). Estimating Variance From High, Low and Closing Prices. *The Annals of Applied Probability*, 1(4):504–512.

Ross, S. A. (1976). The arbitrage theory of capital asset pricing. *Journal of Economic Theory*, 13(3):341–360.

Rzayev, K., Sakkas, A., and Urquhart, A. (2024). An adoption model of cryptocurrencies. *European Journal of Operational Research*.

Sarafidis, V. and Wansbeek, T. (2012). Cross-Sectional Dependence in Panel Data Analysis. *Econometric Reviews*, 31(5):483–531.

Sarafidis, V. and Wansbeek, T. (2021). Celebrating 40 years of panel data analysis: Past, present and future. *Journal of Econometrics*, 220(2):215–226.

Shapley, L. S. (1953). A value for $n$-person games. In Kuhn, H. W. and Tucker, A. W., editors, *Contributions to the Theory of Games II*, volume 28 of *Annals of Mathematics Studies*, pages 307–317. Princeton University Press, Princeton, NJ.

Sharpe, W. F. (1964). Capital asset prices: A theory of market equilibrium under conditions of risk. *The Journal of Finance*, 19(3):425–442.

Sockin, M. and Xiong, W. (2023). A model of cryptocurrencies. *Management Science*, 69:6684–6707.

University of Michigan (2025a). Surveys of Consumers. http://www.sca.isr.umich.edu/. Accessed: 2025-01-21.

University of Michigan (2025b). University of Michigan: Inflation expectation [MICH]. https://fred.stlouisfed.org/series/MICH. retrieved 02/02/2025.

Urquhart, A. (2016). The inefficiency of Bitcoin. *Economics Letters*, 148:80–82.

Wang, Y., Lucey, B. M., Vigne, S. A., and Yarovaya, L. (2022). The effects of central bank digital currencies news on financial markets. *Technological Forecasting and Social Change*, 180:121715.

Yang, D. and Zhang, Q. (2000). Drift-independent volatility estimation based on high, low, open, and close prices. *The Journal of Business*, 73(3):477–91.

Zhang, M., Zhu, B., Li, Z., Jin, S., and Xia, Y. (2024). Relationships among return and liquidity of cryptocurrencies. *Financial Innovation*, 10:3.

**Appendix A. DATA DESCRIPTION**

Table A1 provides a detailed classification of the crypto assets in our sample. Assets that do not use the "Proof of Work" consensus mechanism are considered green crypto assets, as they utilize more energy-efficient protocols such as Proof-of-Stake (PoS), Proof-of-Space-and-Time (PoST), Proof-of-Authority (PoA), and Byzantine Fault Tolerance. Based on this criterion, the green assets in our sample are: ADA, ALGO, ATOM, BNB, CRO, EOS, FIL, IOTA, IRIS, KAVA, RUNE, THETA, XLM, XRP, and XTZ. This classification applies throughout most of the sample period, up to September 15, 2022, when Ethereum transitioned from PoW to PoS.

For the classification of stable versus non-stable assets, we define stablecoins as crypto assets designed to maintain a stable value by pegging their price to a reference asset; typically a fiat currency such as the US dollar or a commodity like gold. The stablecoins in our sample are USDT, USDC, BUSD, DAI, GLC, and PAXG.

Lastly, we identify DeFi assets as those associated with decentralized finance applications, platforms, or protocols that deliver financial services through blockchain technology and smart contracts. Inclusion in the DeFi category is based on an asset's role in supporting DeFi applications, such as enabling staking, tokenization, or liquidity provision. Accordingly, the DeFi assets in our dataset are: ADA, ALGO, ANKR, BNB, CRO, DAI, ETH, FET, FIL, KAVA, MATIC, MKR, PERL, RUNE, USDC, and USDT.

Table A1: Crypto Asset Symbols, Names, Market Capitalizations (USD), and Market Cap Percentages. Source: https://www.coingecko.com/en. Date retrieved: 14/01/2025

| Symbol | Name | Market Cap. | Market Cap. (%) | Classification |
|---|---|---|---|---|
| ADA | Cardano | 3.567E+10 | 1.018 | DeFi, Green |
| ALGO | Algorand | 3.092E+09 | 0.088 | DeFi, Green |
| ANKR | Ankr Network | 3.594E+08 | 0.010 | DeFi |
| ATOM | Cosmos Hub | 2.695E+09 | 0.077 | Green |
| BCH | Bitcoin Cash | 8.659E+09 | 0.247 | |
| BNB | BNB | 1.018E+11 | 2.905 | DeFi, Green |
| BTC | Bitcoin | 1.916E+12 | 54.673 | |
| BTG | Bitcoin Gold | 2.108E+08 | 0.006 | |
| BUSD | BUSD | 5.998E+07 | 0.002 | Stable |
| CRO | Cronos | 3.758E+09 | 0.107 | DeFi, Green |
| DAI | Dai | 3.504E+09 | 0.100 | DeFi, Stable |
| DASH | Dash | 4.496E+08 | 0.013 | |
| DOGE | Dogecoin | 5.233E+10 | 1.493 | |
| EOS | EOS | 1.206E+09 | 0.034 | Green |
| ETC | Ethereum Classic | 3.806E+09 | 0.109 | |
| ETH | Ethereum | 3.880E+11 | 11.074 | DeFi |
| FET | A.S. Alliance† | 3.352E+09 | 0.096 | DeFi |
| FIL | Filecoin | 3.243E+09 | 0.093 | DeFi, Green |
| GLC | Goldcoin | 1.407E+07 | 0.000 | Stable |
| IOTA | IOTA | 1.251E+09 | 0.036 | Green |
| IRIS | IRISnet | 5.418E+06 | 0.000 | Green |
| KAVA | Kava | 5.074E+08 | 0.014 | DeFi, Green |
| LINK | Chainlink | 1.279E+10 | 0.365 | |
| LTC | Litecoin | 7.639E+09 | 0.218 | |
| MANA | Decentraland | 9.171E+08 | 0.026 | |
| MATIC | Polygon | 8.599E+08 | 0.025 | DeFi |
| MKR | Maker | 1.232E+09 | 0.035 | DeFi |
| OMG | OMG Network | 4.381E+07 | 0.001 | |
| PAXG | PAX Gold | 5.332E+08 | 0.015 | Stable |
| PERL | PERL.eco | 5.274E+05 | 0.000 | DeFi |
| RUNE | THORChain | 1.132E+09 | 0.032 | DeFi, Green |
| STX | Stacks | 2.248E+09 | 0.064 | |
| THETA | Theta Network | 2.177E+09 | 0.062 | Green |
| USDC | USDC | 4.569E+10 | 1.304 | DeFi, Stable |
| USDT | Tether | 1.372E+11 | 3.915 | DeFi, Stable |
| XLM | Stellar | 1.303E+10 | 0.372 | Green |
| XMR | Monero | 3.814E+09 | 0.109 | |
| XRP | XRP | 1.533E+11 | 4.374 | Green |
| XTZ | Tezos | 1.305E+09 | 0.037 | Green |
| ZEC | Zcash | 8.115E+08 | 0.023 | |
| | % of market capitalization of the sample | | 83.174 | |

Notes:

† Artificial Superintelligence Alliance.

Table A2: Full Set of Candidate Economy-Wide Predictors

| Name | Short description | Source |
| --- | --- | --- |
| Geopolitical risk index | Measure of adverse geopolitical events based on a tally of newspaper articles covering geopolitical tensions | Caldara and Iacoviello (2022) |
| Global economic policy uncertainty index | GEPU index value is the weighted average (by GDP) of each country's EPU index, based on the share of own-country newspaper articles that discuss economic policy uncertainty each month | Baker et al. (2016) |
| World uncertainty Index | The WUI is computed by counting the percent of word "uncertain" (or its variant) in the Economist Intelligence Unit country reports, and rescaled by multilying by 1,000,000 | Ahir et al. (2022) |
| Macroeconomic and financial uncertainty indices (1-month, 3-month, and 12-month) | Factor analysis based on 132 mostly macro-economic series, and factor analysis 147 financial time series | Jurado et al. (2015) |
| University of Michigan Consumer sentiment | Monthly survey that gauges how confident U.S. consumers feel about their personal finances and the overall economy | University of Michigan (2025a) |
| Policy news and stock market volatility | Newspaper-based equity market volatility (EMV) tracker that moves with the VIX and with the realized volatility of returns on the S&P 500 | Baker et al. (2019) |
| Financial stress index | Measures the degree of financial stress in the markets and is constructed from 18 weekly data series: seven interest rate series, six yield spreads and five other indicators | Federal Reserve Bank of St. Louis (2025) |
| Google Trends cryptocurrency attention | Captures cryptocurrency market attention, based on the rescaled frequency of words related to cryptocurrency market | Aslanidis et al. (2022) |
| University of Michigan inflation expectation (1-Year) | Median expected price change next 12 months based on surveys of consumers | University of Michigan (2025b) |
| Expected inflation (1-Year, 3-Year, 5-Year) | Estimates are calculated with a model that uses Treasury yields, inflation data, inflation swaps, and survey-based measures of inflation expectations | Federal Reserve Bank of Cleveland (2025) |
| Uncertainty of cryptocurrency policy and price (UCRY policy, UCRY price) | Indices based on a query of a set of words ran on LexisNexis database. | Lucey et al. (2022) |
| Climate risk series | Physical and transition risk indices based on text-analysis, using two global climate risk vocabularies | Bua et al. (2024) |
| Climate policy uncertainty index | Index based on searches for articles in eight leading US newspapers containing the terms related to uncertaninty, climate, environment, CO2, and analogous terms | Gavriilidis (2021) |

Table A2 – continued from previous page

| Name | Short description | Source |
| --- | --- | --- |
| Central bank digital currency uncertainty and attention indices | Index based on news stories form LexisNexis News & Business | Wang et al. (2022) |
| Investor sentiment index | Sentiment score obtained by combining six different measures, using the first principal component of the six proxies and their lags | Baker and Wurgler (2006) |
| Chicago Fed national financial condition leverage subindex | Provides a comprehensive weekly update on U.S. financial conditions in money markets, debt and equity markets, and the traditional and "shadow" banking systems, and consists of debt and equity measures | Federal Reserve Bank of Chicago (2025) |
| Monetary policy uncertainty index | Index constructed as scaled frequency counts of newspaper articles that discuss monetary policy uncertainty | Baker et al. (2016) |
| Nominal broad U.S. dollar index | Measures the value of the U.S. dollar against a broad basket of currencies of major U.S. trading partners, weighted by their bilateral trade with the U.S. | Board of Governors of the Federal Reserve System (2025) |

## Appendix B. THEORETICAL RESULTS

**Lemma A.1.** *Under Assumptions 1-7, as $(N,T) \overset{j}{\to} \infty$ such that $N/T \to c$ with $0 < c < \infty$, for $i = 1, 2, .., N$ and $k = 1, 2, .., \ell$,*

$$T^{-r/2} \left\| \widehat{\mathbf{F}} - \mathbf{F}^0 \mathbf{H}_x \right\|^r = T^{-r/2} \sum_{t=1}^{T} \left\| \widehat{\mathbf{f}}_t - \mathbf{H}'_x \mathbf{f}_t^0 \right\|^r = O_p\left(\delta_{NT}^{-r}\right), \, r = 1, 2, \quad (A.1)$$

$$\frac{\left(\widehat{\mathbf{F}} - \mathbf{F}^0 \mathbf{H}_x\right)' \widehat{\mathbf{F}}}{T} = O_p\left(\delta_{NT}^{-2}\right), \quad (A.2)$$

$$\frac{\left(\widehat{\mathbf{F}} - \mathbf{F}^0 \mathbf{H}_x\right)' \mathbf{F}^0}{T} = O_p\left(\delta_{NT}^{-2}\right), \quad (A.3)$$

$$\frac{\left(\widehat{\mathbf{F}} - \mathbf{F}^0 \mathbf{H}_x\right)' \mathbf{G}^0}{T} = O_p\left(\delta_{NT}^{-2}\right), \quad (A.4)$$

$$\frac{\left(\widehat{\mathbf{F}} - \mathbf{F}^0 \mathbf{H}_x\right)' \boldsymbol{\varepsilon}_i}{T} = O_p\left(\delta_{NT}^{-2}\right), \quad (A.5)$$

$$\frac{\left(\widehat{\mathbf{F}} - \mathbf{F}^0 \mathbf{H}_x\right)' \mathbf{v}_{k,i}}{T} = O_p\left(\delta_{NT}^{-2}\right), \quad (A.6)$$

$$\frac{\left(\widehat{\mathbf{F}} - \mathbf{F}^0 \mathbf{H}\right)' \mathbf{C}_i}{T} = O_p\left(\delta_{NT}^{-2}\right), \quad (A.7)$$

$$\frac{1}{\sqrt{N}} \sum_{i=1}^{N} \frac{\left(\widehat{\mathbf{F}} - \mathbf{F}^0 \mathbf{H}_x\right)' \mathbf{v}_{k,i}}{T} \boldsymbol{\gamma}_{k,i}^{0\prime} = O\left(N^{-1/2}\right) + O_p\left(\delta_{NT}^{-2}\right), \quad (A.8)$$

$$\mathbf{H}_x \mathbf{H}'_x - \left(\frac{\mathbf{F}^{0\prime} \mathbf{F}^0}{T}\right)^{-1} = O_p\left(\delta_{NT}^{-2}\right), \quad (A.9)$$

$$\frac{\mathbf{F}^{0\prime} \widehat{\mathbf{F}}}{T} \overset{p}{\to} \boldsymbol{\Lambda} \text{ as } (N,T) \overset{j}{\to} \infty, \quad (A.10)$$

*where $\delta_{NT} = \min\{\sqrt{N}, \sqrt{T}\}$, and we define $\mathbf{F} = \mathbf{F}^0 \mathbf{H}_x$, $\mathbf{G} = \mathbf{G}^0 \mathbf{H}_y$, and $\boldsymbol{\Lambda}_i = \mathbf{H}_x^{-1} \boldsymbol{\Lambda}_i^0$, where $\mathbf{H}_x$ and $\mathbf{H}_y$ are invertible $K_f \times K_f$ and $K_g \times K_g$ matrices.*

*Proof of Lemma A.1.* The proof of (A.1) is given in Bai (2009), Proposition A.1. No modification is required because of our assumption of cross-sectional independence and serial correlation of $v_{k,i,t}$, see Assumption 2. A similar point applies to the proofs of (A.2)-(A.9), which are given by Bai (2009) as proofs of corresponding Lemmas A3(ii), A4(i), A4(ii), A3(iv), A4(iii) and A7(i). The result (A.10) is given as part of Proposition 1 in Bai (2003) with its proof therein. □

**Lemma A.2.** *Under Assumptions 1-7, as $(N,T) \xrightarrow{j} \infty$ such that $N/T \to c$ with $0 < c < \infty$,*

$$T^{-1/2}\mathbf{Z}'_i\mathbf{M_Y}\left(\mathbf{M}_{\widehat{\mathbf{F}}} - \mathbf{M}_{\mathbf{F}^0}\right)\mathbf{u}_i = \sqrt{T}O_p\left(\delta_{NT}^{-2}\right), \tag{A.11}$$

$$T^{-1/2}\mathbf{Z}'_i\mathbf{M_Y}\mathbf{M}_{\widehat{\mathbf{F}}_{-\tau}}\left(\mathbf{M}_{\widehat{\mathbf{F}}} - \mathbf{M}_{\mathbf{F}^0}\right)\mathbf{u}_i = \sqrt{T}O_p\left(\delta_{NT}^{-2}\right), \tag{A.12}$$

$$T^{-1/2}\mathbf{Z}'_i\mathbf{M_Y}\left(\mathbf{M}_{\widehat{\mathbf{F}}_{-\tau}} - \mathbf{M}_{\mathbf{F}^0_{-\tau}}\right)\mathbf{M}_{\mathbf{F}^0}\mathbf{u}_i = \sqrt{T}O_p\left(\delta_{NT}^{-2}\right). \tag{A.13}$$

*Proof of Lemma A.2.* We begin with (A.11). By using $\widehat{\mathbf{F}}'\widehat{\mathbf{F}}/T = \mathbf{I}_{K_f}$, we have $\mathbf{M}_{\widehat{\mathbf{F}}} - \mathbf{M}_{F^0} = \mathbf{P}_{\mathbf{F}^0} - \mathbf{P}_{\widehat{\mathbf{F}}} = -\left(\frac{\widehat{\mathbf{F}}\widehat{\mathbf{F}}'}{T} - \mathbf{P}_{\mathbf{F}^0}\right)$. Therefore, we get

$$\begin{aligned}
&T^{-1/2}\mathbf{Z}'_i\mathbf{M_Y}(\mathbf{M}_{\widehat{\mathbf{F}}} - \mathbf{M}_{\mathbf{F}^0})\mathbf{u}_i \\
&= T^{-1/2}\mathbf{Z}'_i\mathbf{M_Y}(\mathbf{M}_{\widehat{\mathbf{F}}} - \mathbf{M}_{\mathbf{F}^0})\mathbf{u}_i \\
&= -\frac{1}{\sqrt{T}}\mathbf{Z}'_i\mathbf{M_Y}\left(\frac{\widehat{\mathbf{F}}\widehat{\mathbf{F}}'}{T} - \mathbf{P}_{\mathbf{F}^0}\right)\mathbf{u}_i \\
&= -\frac{1}{\sqrt{T}}\frac{\mathbf{Z}'_i\mathbf{M_Y}\left(\widehat{\mathbf{F}} - \mathbf{F}^0\mathbf{H}_x\right)}{T}\mathbf{H}'_x\mathbf{F}^{0'}\mathbf{u}_i \\
&\quad -\frac{1}{\sqrt{T}}\frac{\mathbf{Z}'_i\mathbf{M_Y}\left(\widehat{\mathbf{F}} - \mathbf{F}^0\mathbf{H}_x\right)}{T}\left(\widehat{\mathbf{F}} - \mathbf{F}^0_x\mathbf{H}_x\right)'\mathbf{u}_i \\
&\quad -\frac{1}{\sqrt{T}}\frac{\mathbf{Z}'_i\mathbf{M_Y}\mathbf{F}^0}{T}\mathbf{H}_x\left(\widehat{\mathbf{F}} - \mathbf{F}^0\mathbf{H}_x\right)'\mathbf{u}_i \\
&\quad -\frac{1}{\sqrt{T}}\frac{\mathbf{Z}'_i\mathbf{M_Y}\mathbf{F}^0}{T}\left[\mathbf{H}_x\mathbf{H}'_x - \left(\frac{\mathbf{F}^{0'}\mathbf{F}^0_x}{T}\right)^{-1}\right]\mathbf{F}^{0'}\mathbf{u}_i \\
&= -(\mathbf{a}_1 + \mathbf{a}_2 + \mathbf{a}_3 + \mathbf{a}_4),
\end{aligned}$$

$$\begin{aligned}
|\mathbf{a}_1| &\leq \sqrt{T}\left\|\frac{\mathbf{Z}'_i\mathbf{M_Y}\left(\widehat{\mathbf{F}} - \mathbf{F}^0\mathbf{H}_x\right)}{T}\right\|\|\mathbf{H}_x\|\left\|\frac{\mathbf{F}^{0'}\mathbf{u}_i}{T}\right\| \\
&\leq \sqrt{T}\|\mathbf{\Lambda}_i\|\left\|\frac{\mathbf{F}'\left(\widehat{\mathbf{F}} - \mathbf{F}^0\mathbf{H}_x\right)}{T}\right\|\|\mathbf{H}_x\|\left\|\frac{\mathbf{F}^0}{\sqrt{T}}\right\|\left\|\frac{\mathbf{u}_i}{\sqrt{T}}\right\| \\
&\quad + \sqrt{T}\left\|\frac{\mathbf{V}'_i\left(\widehat{\mathbf{F}} - \mathbf{F}^0\mathbf{H}_x\right)}{T}\right\|\|\mathbf{H}_x\|\left\|\frac{\mathbf{F}^0}{\sqrt{T}}\right\|\left\|\frac{\mathbf{u}_i}{\sqrt{T}}\right\| \\
&= \sqrt{T}O_p\left(\delta_{NT}^{-2}\right),
\end{aligned}$$

from $\left\|\frac{\mathbf{F}^{0'}(\widehat{\mathbf{F}}-\mathbf{F}^0\mathbf{H}_x)}{T}\right\| = O_p\left(\delta_{NT}^{-2}\right)$ and $\left\|\frac{\mathbf{V}'_i(\widehat{\mathbf{F}}-\mathbf{F}^0\mathbf{H}_x)}{T}\right\| = O_p\left(\delta_{NT}^{-2}\right)$ by Lemma A.1, $\|\mathbf{H}_x\| = $

$O_p(1)$, $\|\mathbf{\Lambda}_i\| = O_p(1)$ by Assumption 4, $\frac{\|\mathbf{F}^0\|}{\sqrt{T}} = O_p(1)$, and $\frac{\|\mathbf{u}_i\|}{\sqrt{T}} \leq \|\boldsymbol{\gamma}_i^0\| \frac{\|\mathbf{G}^0\|}{\sqrt{T}} + \|\boldsymbol{\gamma}_i^0\| \frac{\|\mathbf{G}^0\|}{\sqrt{T}} + \frac{\|\boldsymbol{\varepsilon}_i\|}{\sqrt{T}} = O_p(1)$ by Assumptions 1, 3 and 4.

$$|\mathbf{a}_2| \leq \sqrt{T} \left\| \frac{\mathbf{Z}_i' \mathbf{M_Y} (\widehat{\mathbf{F}} - \mathbf{F}^0 \mathbf{H}_x)}{T} \right\| \left\| \frac{(\widehat{\mathbf{F}} - \mathbf{F}^0 \mathbf{H}_x)' \mathbf{u}_i}{T} \right\|$$

$$\leq \sqrt{T} \|\mathbf{\Lambda}_i\| \left\| \frac{\mathbf{F}' (\widehat{\mathbf{F}} - \mathbf{F}^0 \mathbf{H}_x)}{T} \right\| \left\| \frac{(\widehat{\mathbf{F}} - \mathbf{F}^0 \mathbf{H}_x)' \mathbf{u}_i}{T} \right\|$$

$$+ \sqrt{T} \left\| \frac{\mathbf{V}_i' (\widehat{\mathbf{F}} - \mathbf{F}^0 \mathbf{H}_x)}{T} \right\| \left\| \frac{(\widehat{\mathbf{F}} - \mathbf{F}^0 \mathbf{H}_x)' \mathbf{u}_i}{T} \right\|$$

$$= \sqrt{T} O_p(\delta_{NT}^{-4}),$$

by similar arguments as above and $\left\| \frac{\mathbf{u}_i' (\widehat{\mathbf{F}} - \mathbf{F}^0 \mathbf{H}_x)}{T} \right\| \leq \|\boldsymbol{\lambda}_i^0\| \left\| \frac{\mathbf{F}^{0\prime} (\widehat{\mathbf{F}} - \mathbf{F}^0 \mathbf{H}_x)}{T} \right\| + \|\boldsymbol{\delta}_i^0\| \left\| \frac{\mathbf{G}^{0\prime} (\widehat{\mathbf{F}} - \mathbf{F}^0 \mathbf{H}_x)}{T} \right\| + \left\| \frac{\boldsymbol{\varepsilon}_i' (\widehat{\mathbf{F}} - \mathbf{F}^0 \mathbf{H}_x)}{T} \right\| = O_p(\delta_{NT}^{-2})$ by Lemma A.1 and Assumption 4.

$$|\mathbf{a}_3| \leq \sqrt{T} \left\| \frac{\mathbf{Z}_i' \mathbf{M_Y} \mathbf{F}_x^0}{T} \right\| \|\mathbf{G}\| \left\| \frac{(\widehat{\mathbf{F}}_x - \mathbf{F}_x^0 \mathbf{G})' \mathbf{u}_i}{T} \right\|$$

$$\leq \sqrt{T} \|\mathbf{\Lambda}_i\| \left\| \frac{\mathbf{F}^0}{\sqrt{T}} \right\|^2 \|\mathbf{H}_x\| \left\| \frac{(\widehat{\mathbf{F}} - \mathbf{F}^0 \mathbf{H}_x)' \mathbf{u}_i}{T} \right\|$$

$$+ \sqrt{T} \left\| \frac{\mathbf{V}_i}{\sqrt{T}} \right\| \left\| \frac{\mathbf{F}^0}{\sqrt{T}} \right\| \|\mathbf{H}_x\| \left\| \frac{(\widehat{\mathbf{F}} - \mathbf{F}^0 \mathbf{H}_x)' \mathbf{u}_i}{T} \right\|$$

$$= \sqrt{T} O_p(\delta_{NT}^{-2}),$$

from $\left\| \frac{\mathbf{u}_i' (\widehat{\mathbf{F}} - \mathbf{F}^0 \mathbf{H}_x)}{T} \right\| = O_p(\delta_{NT}^{-2})$ as shown above by using Lemma A.1 and Assumption 4, $\|\mathbf{H}_x\| = O_p(1)$, $\|\mathbf{\Lambda}_i\| = O_p(1)$ by Assumption 4 as well as $\frac{\|\mathbf{F}^0\|}{\sqrt{T}} = O_p(1)$ by Assumption 3.

$$|\mathbf{a}_4| \leq \sqrt{T} \left\| \frac{\mathbf{Z}_i' \mathbf{M_Y F}^0}{T} \right\| \left\| \mathbf{H}_x \mathbf{H}_x' - \left( \frac{\mathbf{F}^{0\prime} \mathbf{F}^0}{T} \right)^{-1} \right\| \left\| \frac{\mathbf{F}^{0\prime} \mathbf{u}_i}{T} \right\|$$

$$\leq \sqrt{T} \|\mathbf{\Lambda}_i\| \left\| \frac{\mathbf{F}^0}{\sqrt{T}} \right\|^2 \left\| \mathbf{H}_x \mathbf{H}_x' - \left( \frac{\mathbf{F}^{0\prime} \mathbf{F}^0}{T} \right)^{-1} \right\| \left\| \frac{\mathbf{u}_i}{\sqrt{T}} \right\|$$

$$+ \sqrt{T} \left\| \frac{\mathbf{V}_i}{\sqrt{T}} \right\| \left\| \frac{\mathbf{F}^0}{\sqrt{T}} \right\| \left\| \mathbf{H}_x \mathbf{H}_x' - \left( \frac{\mathbf{F}^{0\prime} \mathbf{F}^0}{T} \right)^{-1} \right\| \left\| \frac{\mathbf{u}_i}{\sqrt{T}} \right\|$$

$$= \sqrt{T} O_p \left( \delta_{NT}^{-2} \right).$$

from making use of $\left\| \mathbf{H}_x \mathbf{H}_x' - (T^{-1} \mathbf{F}^{0\prime} \mathbf{F}^0)^{-1} \right\| = O_p \left( \delta_{NT}^{-2} \right)$ by Lemma A.1, $\frac{\|\mathbf{V}_i\|}{\sqrt{T}} = O_p(1)$ by Assumption 2, $\frac{\|\mathbf{F}^0\|}{\sqrt{T}} = O_p(1)$ by Assumption 3, $\|\mathbf{\Lambda}_i\| = O_p(1)$ by Assumption 4 and $\frac{\|\mathbf{u}_i\|}{\sqrt{T}} = O_p(1)$ by Assumptions 1, 3 and 4.

By putting the results together, we therefore have

$$\left\| T^{-1/2} \mathbf{Z}_i' \mathbf{M_Y} (\mathbf{M_{\widehat{F}}} - \mathbf{M_{F^0}}) \mathbf{u}_i \right\| = \sqrt{T} O_p \left( \delta_{NT}^{-2} \right). \tag{A.14}$$

Thus,

$$T^{-1/2} \mathbf{Z}_i' \mathbf{M_Y} \mathbf{M_{\widehat{F}}} \mathbf{u}_i = T^{-1/2} \mathbf{Z}_i' \mathbf{M_Y} \mathbf{M_{F^0}} \mathbf{u}_i + \sqrt{T} O_p \left( \delta_{NT}^{-2} \right). \tag{A.15}$$

The results in (A.12) and (A.13) can be shown in a similar manner. $\square$

The following proposition establishes the limiting property of $T^{-1/2} \widehat{\mathbf{Z}}_i' \mathbf{M_{\widehat{F}}} \mathbf{u}_i$, which is necessary for Theorem 1.

**Proposition A.1.** *Consider the model in Eq. (1). Under Assumptions 1–7, as $(N,T) \xrightarrow{j} \infty$ such that $N/T \to c$ with $0 < c < \infty$, we have*

$$T^{-1/2} \widehat{\mathbf{Z}}_i' \mathbf{M_{\widehat{F}}} \mathbf{u}_i = T^{-1/2} \mathbf{Z}_i' \mathbf{M_Y} \mathbf{M_{F^0}} \mathbf{u}_i + \sqrt{T} O_p \left( \delta_{NT}^{-2} \right),$$

*where $\delta_{NT} = \min \left\{ \sqrt{T}, \sqrt{N} \right\}$.*

*Proof of Proposition A.1.* Consider $T^{-1/2} \widehat{\mathbf{Z}}_i' \mathbf{M_{\widehat{F}}} \mathbf{u}_i$,
where $\widehat{\mathbf{Z}}_i = \left( \mathbf{M_{\widehat{F}}} \mathbf{M_Y} \mathbf{Z}_i, \ldots, \mathbf{M_{\widehat{F}_{-\zeta}}} \mathbf{M_Y} \mathbf{Z}_i, \mathbf{Y}, \mathbf{Y}_{,-1} \right)$.

Let us start with the first component of $\mathbf{M_{\widehat{F}}} \widehat{\mathbf{Z}}_i$, i.e. $\mathbf{M_{\widehat{F}}} \mathbf{M_Y} \mathbf{Z}_i$. By adding and subtracting we get

$$T^{-1/2} \mathbf{Z}_i' \mathbf{M_Y} \mathbf{M_{\widehat{F}}} \mathbf{u}_i = T^{-1/2} \mathbf{Z}_i' \mathbf{M_Y} \mathbf{M_{F^0}} \mathbf{u}_i + T^{-1/2} \mathbf{Z}_i' \mathbf{M_Y} \left( \mathbf{M_{\widehat{F}}} - \mathbf{M_{F^0}} \right) \mathbf{u}_i$$

$$= T^{-1/2} \mathbf{Z}_i' \mathbf{M_Y} \mathbf{M_{F^0}} \mathbf{u}_i + \sqrt{T} O_p \left( \delta_{NT}^{-2} \right) \tag{A.16}$$

where the second equality is due to result in (A.11) stated in Lemma A.2.

Next is the second component of $\mathbf{M}_{\widehat{\mathbf{F}}}\widehat{\mathbf{Z}}_i$, which is $\mathbf{M}_{\widehat{\mathbf{F}}}\mathbf{M}_{\widehat{\mathbf{F}}_{-\tau}}\mathbf{M}_{\mathbf{Y}}\mathbf{Z}_{i,-\tau}$. Again, by adding and subtracting and using Lemma A.2, we get

$$
\begin{aligned}
&T^{-1/2}\mathbf{Z}'_{i,-\tau}\mathbf{M}_{\mathbf{Y}}\mathbf{M}_{\widehat{\mathbf{F}}_{-\tau}}\mathbf{M}_{\widehat{\mathbf{F}}}\mathbf{u}_i \\
&= T^{-1/2}\mathbf{Z}'_{i,-1}\mathbf{M}_{\mathbf{Y}}\mathbf{M}_{\widehat{\mathbf{F}}_{-\tau}}\mathbf{M}_{\mathbf{F}^0}\mathbf{u}_i + T^{-1/2}\mathbf{Z}'_i\mathbf{M}_{\mathbf{Y}}\mathbf{M}_{\widehat{\mathbf{F}}_{-\tau}}\left(\mathbf{M}_{\widehat{\mathbf{F}}} - \mathbf{M}_{\mathbf{F}^0}\right)\mathbf{u}_i \\
&= T^{-1/2}\mathbf{Z}'_{i,-1}\mathbf{M}_{\mathbf{Y}}\mathbf{M}_{\widehat{\mathbf{F}}_{-\tau}}\mathbf{M}_{\mathbf{F}^0}\mathbf{u}_i + \sqrt{T}O_p\left(\delta_{NT}^{-2}\right) \\
&= T^{-1/2}\mathbf{Z}'_{i,-1}\mathbf{M}_{\mathbf{Y}}\mathbf{M}_{\mathbf{F}^0_{-\tau}}\mathbf{M}_{\mathbf{F}^0}\mathbf{u}_i + T^{-1/2}\mathbf{Z}'_{i,-\tau}\mathbf{M}_{\mathbf{Y}}\left(\mathbf{M}_{\widehat{\mathbf{F}}_{-\tau}} - \mathbf{M}_{\mathbf{F}^0_{-\tau}}\right)\mathbf{M}_{\mathbf{F}^0}\mathbf{u}_i + \sqrt{T}O_p\left(\delta_{NT}^{-2}\right) \\
&= T^{-1/2}\mathbf{Z}'_{i,-1}\mathbf{M}_{\mathbf{Y}}\mathbf{M}_{\mathbf{F}^0_{-\tau}}\mathbf{M}_{\mathbf{F}^0_x}\mathbf{u}_i + \sqrt{T}O_p\left(\delta_{NT}^{-2}\right).
\end{aligned} \quad (A.17)
$$

Finally, by combining the results, we get

$$T^{-1/2}\widehat{\mathbf{Z}}'_i\mathbf{M}_{\widehat{\mathbf{F}}}\mathbf{u}_i = T^{-1/2}\mathbf{Z}'_i\mathbf{M}_{\mathbf{Y}}\mathbf{M}_{\mathbf{F}^0}\mathbf{u}_i + \sqrt{T}O_p\left(\delta_{NT}^{-2}\right), \quad (A.18)$$

where $\mathbf{Z}_i = \left(\mathbf{M}_{\mathbf{F}^0}\mathbf{M}_{\mathbf{Y}}\mathbf{Z}_i, \ldots, \mathbf{M}_{\mathbf{F}^0_{-\zeta}}\mathbf{M}_{\mathbf{Y}}\mathbf{Z}_i, \mathbf{Y}, \mathbf{Y}_{,-1}\right)$, which provides the expression given in Proposition A.1. $\square$

*Proof of Theorem 1.* By Proposition 3 $T^{-1/2}\widehat{\mathbf{Z}}'_i\mathbf{M}_{\widehat{\mathbf{F}}}\mathbf{u}_i = T^{-1/2}\mathbf{Z}'_i\mathbf{M}_{\mathbf{Y}}\mathbf{M}_{\mathbf{F}^0}\mathbf{u}_i + o_p(1)$ as $(N,T) \xrightarrow{j} \infty$ as $N/T \to c$ for $0 < c < \infty$. It is immediate that, under Assumptions 1-7, for each $i$, $T^{-1/2}\mathbf{Z}'_i\mathbf{M}_{\mathbf{Y}}\mathbf{u}_i \xrightarrow{d} N(\mathbf{0}, \boldsymbol{\Sigma}_i)$. Similar lines of argument ensure that $\widehat{\mathbf{A}}_{i,T} - \mathbf{A}_{i,T} \xrightarrow{p} \mathbf{0}$ and $\widehat{\mathbf{B}}_{i,T} - \mathbf{B}_{i,T} \xrightarrow{p} \mathbf{0}$ as $T \to \infty$, and together with Assumption 7 we see that $p\lim_{T\to\infty}\widehat{\mathbf{A}}_{i,T} = \mathbf{A}_i$ and $p\lim_{T\to\infty}\widehat{\mathbf{B}}_{i,T} = \mathbf{B}_i$, thus the required result follows. $\square$

*Proof of Proposition 1.* Assume that the IV residuals admit a factor structure of the form:

$$u_{i,t} = \boldsymbol{\delta}'_i\mathbf{g}_t + \varepsilon_{i,t}, \quad i = 1,\ldots,N, \quad t = 1,\ldots,T, \quad (A.19)$$

where $\mathbf{g}_t \in \mathbb{R}^{K_g}$ denotes a vector of latent common factors, $\boldsymbol{\delta}_i \in \mathbb{R}^{K_g}$ are unit-specific loadings, and $\varepsilon_{i,t}$ is an idiosyncratic error with zero mean and finite variance. In vector form, we have

$$\mathbf{u}_i = \mathbf{G}\boldsymbol{\delta}'_i + \boldsymbol{\varepsilon}_i, \quad (A.20)$$

where $\mathbf{G} = (\mathbf{g}_1, \ldots, \mathbf{g}_T)'$ is $T \times K_g$.

Let

$$\mathbf{F} = \mathbf{G}\mathbf{H}, \quad (A.21)$$

where $\mathbf{H} \in \mathbb{R}^{\widehat{K}_g \times K_g}$ is a full-rank rotation matrix that maps the latent factor space into an identifiable rotated representation. We assume that the first row of $\mathbf{H}$ contains no zero entries – a standard and mild condition in the approximate factor model literature. This ensures that the leading principal component is a nontrivial linear combination of all latent factors.[14]

---

[14]If this assumption is not tenable, one may extend the procedure by applying the MTB algorithm separately to each principal component found to be statistically significant. The selected predictors from each

Given this assumption, the probability limit of the leading principal component, denoted $\mathbf{f}_1$, lies in the column space of $\mathbf{G}$ and thus preserves sufficient information to recover the entire latent factor structure.

Next, by the theoretical results in Kapetanios et al. (2025), regressing $\mathbf{f}_1$ on the full set of candidate proxies yields consistent identification of the true factors with probability approaching one.

Finally, invoking Theorem 1 and results in Bai and Ng (2006), the two approximation errors arising by (i) using the estimated residuals $\widehat{\mathbf{u}}_i$ in place of the true errors $\mathbf{u}_i$ when extracting the leading principal component; and (ii) substituting the sample-based $\mathbf{e}_1$ for the population quantity $\mathbf{f}_1$, are asymptotically negligible for the validity of PCA-MTB. □

## Appendix C. MONTE CARLO STUDY

This section evaluates the finite-sample performance of the PCA-MTB method using diagnostic metrics commonly used in high-dimensional variable selection. These metrics are formally defined below.

*Appendix C.1. Design*

The data-generating process (DGP) is given by

$$u_{i,t} = \boldsymbol{\delta}'_i \mathbf{g}_t + \varepsilon_{i,t}, \tag{A.1}$$

where $\mathbf{g}_t$ denotes an $r \times 1$ vector of common factors (or "signals") affecting all individuals albeit with heterogeneous loadings $\boldsymbol{\delta}_i$. The outcome $u_{i,t}$ corresponds to the residual $\widehat{u}_{i,t}$ in the main text, which is $\sqrt{T}$-consistent. The idiosyncratic errors $\varepsilon_{i,t}$ are drawn as $i.i.d. \mathcal{N}(0,1)$ for all $i$ and $t$.

The factors are drawn as

$$\mathbf{g}_t = \rho \mathbf{g}_{t-1} + \sqrt{(1-\rho^2)} \boldsymbol{\epsilon}_t, \tag{A.2}$$

where $\boldsymbol{\epsilon}_t \sim \mathcal{N}_r(\mathbf{0}_r, \mathbf{I}_r)$ and $\mathbf{I}_r$ denotes the $r \times r$ identify matrix. The factor loadings are drawn from

$$\boldsymbol{\delta}_i \sim \mathcal{N}_r(\boldsymbol{\mu}_r, \boldsymbol{\Sigma}_r), \tag{A.3}$$

with $\boldsymbol{\mu}_r = \mathbf{1}'_r \phi$, where $\mathbf{1}_r = (1, 1, \ldots, 1)'$. The covariance matrix $\boldsymbol{\Sigma}_r \in \mathbb{R}^{r \times r}$ is given by

$$\boldsymbol{\Sigma}_r = 0.5 \mathbf{1}_r \mathbf{1}'_r + 0.5 \cdot \mathbf{I}_r, \tag{A.4}$$

so that all diagonal entries equal 1 and all off-diagonal entries equal 0.5, implying substantial mutual correlation among the loadings.

---

component-specific regression can then be combined by taking the union of all selected variables across the $\widehat{R}$ regressions, where $\widehat{R}$ denotes the estimated number of relevant principal components. The latter can be determined using information criteria such as those proposed by Ahn and Horenstein (2013). This method of estimating the number of principal components has been shown to perform well in finite samples across various settings, including dynamic panel models with lagged risk effects, as considered in the present context; see Juodis and Sarafidis (2018), Juodis and Sarafidis (2022b) for further discussion.

In addition to the factors entering in Eq. (A.1), we generate $n$ further common factors that do not enter the DGP in Eq. (A.1):

$$f_{j,t} = \begin{cases} \rho f_{j,t-1} + \sqrt{1-\rho^2}\left(\sqrt{\pi}v_{j,t} + \sqrt{(1-\pi)}\epsilon_{j,t}\right), & \text{if } j \leq r, \\ \rho f_{j,t-1} + \sqrt{1-\rho^2}v_{j,t}, & \text{if } j > r, \end{cases} \quad (A.5)$$

where $v_{j,t} \sim i.i.d.N(0,1)$ and is independent of $\varepsilon_{i,t}$. The first $r$ factors in Eq. (A.5) are referred to as "pseudo-signals" because, although irrelevant, they are correlated with the true factors and may be mistakenly identified as informative. The remaining $n-r$ are pure noise variables. This structure allows us to evaluate the performance of variable selection methods in the presence of confounding and irrelevant variables.

We consider two levels of signal dimensionality, $r \in \{2, 5\}$, with $\rho = 0.5$, $\pi = 0.5$, allowing for sizable persistence and correlation between signals and pseudo-signals. The location parameter $\phi \in \{0, 1\}$ determines whether factor loadings are centered at zero or not, affecting the performance of selection methods that rely on pooling across $i$. We set $N = T$ and $n = T - r$, and explore all combinations of $T \in \{50, 100, 200\}$ and the corresponding values of $n$. This setup enables an assessment of how performance varies with dimensionality across different values $T, n$.[15] For each choice of $r$, the analysis covers a total of 18 configurations.

*Appendix C.2. The Multiple Testing Boosting Method*

The Multiple Testing Boosting (MTB) approach of Kapetanios et al. (2025) is a forward selection procedure designed to recover relevant predictors from a high-dimensional set. It operates iteratively by testing one covariate at a time, updating the model step-by-step, and dynamically adjusting the selection threshold to control the family-wise error rate. In what follows, we provide a high-level description of the method.

Let $\mathbf{e}_1 \in \mathbb{R}^{T \times 1}$ denote the estimated signal component of the data, obtained by extracting the dominant principal component from the sample variance-covariance matrix of $y_{i,t}$. The goal is to identify relevant predictors from a pool of $n$ candidate variables, stored in matrix $\mathbf{Z} = (\mathbf{G}, \mathbf{F}) \in \mathbb{R}^{T \times r+n}$, where $\mathbf{G} = (\mathbf{g}_1, \ldots, \mathbf{g}_T)'$ and $\mathbf{F} = (\mathbf{f}_1, \ldots, \mathbf{f}_T)'$ with $\mathbf{f}_t = (f_{1,t}, \ldots, f_{n,t})'$, such that $\mathbf{Z}_j$ significantly explains variation in $\mathbf{e}_1$.

The algorithm proceeds in iterative passes. At the first pass, a univariate regression is estimated for each predictor:

$$\mathbf{e}_1 = \mathbf{Z}_j \cdot b_j + \text{residuals}, \quad (A.6)$$

and a robust $t$-statistic is computed for the coefficient. The selection threshold accounts for multiple testing via a Bonferroni-type correction:

$$p_j^{(1)} = \frac{p_{\text{val}}}{c_1(n-\varrho)^{\delta_1 - 1}},$$

with $p_{\text{val}} = 0.05$, $c_1 = 1$, $\varrho = 0$, and $\delta_1 = 1.5$.[16] The predictor with the largest absolute $t$-statistic that exceeds this threshold is selected and added to the model.

---

[15] The restriction $N = T$ is motivated by the fact that, in our dataset, the number of crypto assets (40) is of comparable magnitude to the number of monthly observations (36).

[16] The choices for $c_1$ and $\varrho$ are based on Kapetanios et al. (2025).

In subsequent passes, all previously selected predictors are retained as controls. Specifically, at step $k$, the following regression is estimated for each remaining candidate $j \notin \mathcal{I}_{k-1}$:

$$\mathbf{e}_1 = \mathbf{Z}_{\mathcal{I}_{k-1}} \cdot \mathbf{b} + \mathbf{Z}_j \cdot b_j + \text{residuals}, \tag{A.7}$$

where $\mathcal{I}_{k-1}$ denotes the indices selected in the first $k-1$ steps. The robust $t$-statistic for $b_j$ is computed and compared against the updated threshold

$$p_j^{(k)} = \frac{p_{\text{val}}}{c_1(n-\varrho)^{\delta_1-1}}, \quad \text{with } \varrho = k-1.$$

If the largest (absolute) $t$-statistic exceeds the threshold, the corresponding predictor is added to the model. Otherwise, the procedure terminates.

The final output is the set of selected predictors:

$$\widehat{\mathcal{S}} = \{j_1, j_2, \ldots, j_K\},$$

where each $j_k$ denotes the index selected at iteration $k$. Each factor may be selected at most once, and once included, it remains in the model throughout the remaining steps. The number of selected variables is $|\widehat{\mathcal{S}}| = K$.

This iterative framework enables the identification of relevant predictors while mitigating the multiple testing burden through controlled stepwise inference. It is particularly effective when the number of relevant variables is moderately small relatively to the size of $T$, and remains valid even when factors are correlated with irrelevant covariates.

In addition to the combined PCA-MTB approach discussed above, we consider two alternative approaches: a pooled Lasso estimator that imposes homogeneous coefficients across units, and an individual-level Lasso estimator that allows for heterogeneous coefficients.

The pooled Lasso ("p-Lasso", hereafter) estimator solves the following optimization problem:

$$\widehat{\boldsymbol{\beta}}^{\text{p-Lasso}} = \arg\min_{\boldsymbol{\beta} \in \mathbb{R}^{r+n}} \left\{ \frac{1}{NT} \|\mathbf{u}_i - \boldsymbol{\mathcal{Z}}\mathbf{b}\|_2^2 + \xi \sum_{j=1}^{n} |b_j| \right\}, \tag{A.8}$$

where $\xi \geq 0$ is a tuning parameter selected via 10-fold cross-validation. Although this model is misspecified under the DGP (which assumes heterogeneous coefficients $\boldsymbol{\beta}_i = (\boldsymbol{\delta}_i', \mathbf{0}_n')'$), the pooled estimator $\widehat{\boldsymbol{\beta}}$ may still be effective for variable selection under sparsity, particularly when the individual-specific loadings are not centered around zero, i.e., $\phi \neq 0$. It can be interpreted as a regularized projection of the cross-sectional signal onto the predictor space spanned by $\mathbf{z}_t$.

The individual-level Lasso ("i-Lasso", hereafter) estimator instead estimates a separate coefficient vector for each unit $i = 1, \ldots, N$:

$$\widehat{\boldsymbol{\beta}}_i^{\text{i-Lasso}} = \arg\min_{\boldsymbol{\beta}_i \in \mathbb{R}^{r+n}} \left\{ \frac{1}{2T} \|\mathbf{u}_i - \boldsymbol{\mathcal{Z}}\mathbf{b}_i\|_2^2 + \xi_i \sum_{j=1}^{n} |b_{i,j}| \right\}, \tag{A.9}$$

where $\mathbf{b}_i$ is penalized using an individual-specific regularization parameter $\xi_i$, selected via 10-fold cross-validation. To identify relevant factors, we apply a stability selection rule: for each factor $j$, we count how often it is selected across the $N$ individual Lasso estimates (i.e., when $\widehat{\beta}_{i,j} \neq 0$). Predictors that appear in at least 25% of units are retained. The final model size therefore reflects the number of variables that meet this frequency threshold, providing a measure of their cross-sectional relevance.[17]

*Appendix C.3. Results*

To assess the variable selection performance of PCA-MTB and alternative methods, we employ a comprehensive set of diagnostic metrics tailored to high-dimensional selection tasks, as detailed in Kapetanios et al. (2025). These include the True Positive Rate (TPR), False Positive Rate (FPR), True Discovery Rate (TDR), False Discovery Rate (FDR), the F1 score, and the Matthews Correlation Coefficient (MCC). These metrics are computed based on the confusion matrix that categorizes selected variables as true or false discoveries.

The TPR (or recall) measures the proportion of true signals correctly identified, while the TDR (or precision) reflects the proportion of selected variables that are indeed true signals. FPR and FDR capture the share of irrelevant variables mistakenly selected.

The F1 score, defined as the harmonic mean of TPR and TDR, summarizes the trade-off between precision and recall. The MCC offers a more comprehensive measure, incorporating all four components of the confusion matrix and accounting for both Type I and Type II errors. It is particularly valuable in imbalanced settings like ours, where the number of true signals is small relative to the total number of candidates.

We also report the model size, defined as the number of selected variables, as a measure of sparsity. Further details on the computation and interpretation of these metrics are provided in Kapetanios et al. (2025).

We focus primarily on the MCC criterion. Tables A3–A4 present summary statistics across 18 designs for each sparsity level ($r = 2$ and $r = 5$).

PCA-MTB performs strongly, achieving the highest median MCC in both settings (0.958 for $r = 2$, 0.922 for $r = 5$), with relatively narrow interquartile ranges and strong ranking outcomes. Its MCC exceeds the 0.8 threshold in all cases for $r = 2$, and in approximately 78% of cases for $r = 5$. As a result, it ranks first in the vast majority of configurations. Moreover, in the few instances where it ranks second, the difference in MCC compared to the top-ranked method is negligible, as indicated in Table A5.

p-Lasso performs reasonably well in some instances (maximum MCC up to 0.854), but its results are more variable. For $r = 2$, it has a median MCC of 0.713 and a wider IQR (0.228). Further, it does not rank first in any configuration. As discussed further below, its performance appears sensitive to specific design features, particularly whether $\phi$, the population average of factor loadings, equals zero or not.

i-Lasso outperforms p-Lasso, with median MCC values of 0.871 ($r = 2$) and 0.779 ($r = 5$). However, it generally falls short of PCA-MTB in both accuracy and stability across designs.

To gain further insight into method performance across several $T, n$ combinations, Figure C.1 presents a visual summary of MCC-based performance under $r = 2, 5$ and $\phi = 0, 1$. Details results are reported in Tables A5–A11.

---

[17]Naturally, the performance of i-Lasso can be sensitive to the choice of the stability selection threshold. While exploring its optimal calibration (e.g., via cross-validation) might improve results in small samples, such an investigation lies beyond the scope of this paper.

Table A3: Summary Statistics for Matthews Correlation Coefficient (MCC) with $r = 2$

|         | median | IQR   | min   | max   | prop. $> 0.8$ | rank  | prop. 1st |
|---------|--------|-------|-------|-------|---------------|-------|-----------|
| PCA-MTB | 0.958  | 0.045 | 0.878 | 0.979 | 100.0%        | 1.056 | 94.44%    |
| p-Lasso | 0.713  | 0.228 | 0.403 | 0.851 | 27.8%         | 2.833 | 0.00%     |
| i-Lasso | 0.871  | 0.254 | 0.529 | 0.977 | 66.7%         | 2.111 | 5.56%     |

**Notes**: MCC ranges from $-1$ (perfect misclassification) to $+1$ (perfect selection), with 0 indicating random guessing. "rank" refers to the average position of each method when ranked by MCC across all configurations. "prop. $> 0.8$" reports the proportion of cases across all 18 designs in which MCC exceeds 0.8. "prop. 1st" reports the proportion of cases in which a method ranks first based on MCC performance.

Table A4: Summary Statistics for Matthews Correlation Coefficient (MCC) with $r = 5$

|         | median | IQR   | min   | max   | prop. $> 0.8$ | rank  | prop. 1st |
|---------|--------|-------|-------|-------|---------------|-------|-----------|
| PCA-MTB | 0.922  | 0.109 | 0.683 | 0.970 | 77.8%         | 1.111 | 88.89%    |
| p-Lasso | 0.705  | 0.201 | 0.360 | 0.854 | 16.7%         | 2.667 | 0.00%     |
| i-Lasso | 0.779  | 0.235 | 0.587 | 0.943 | 50.0%         | 2.222 | 11.11%    |

**Notes**: MCC ranges from $-1$ (perfect misclassification) to $+1$ (perfect selection), with 0 indicating random guessing. "rank" refers to the average position of each method when ranked by MCC across all configurations. "prop. $> 0.8$" reports the proportion of cases across all 18 designs in which MCC exceeds 0.8. "prop. 1st" reports the proportion of cases in which a method ranks first based on MCC performance.

The graphical results reveal clear and systematic differences in performance across methods. For $r = 2$ and $\phi = 1$, PCA-MTB consistently achieves the highest MCC values across all nine configurations. Its performance improves steadily with larger $T$ or smaller $n$, rising from 0.901 in the smallest panel $(25, 100)$ to 0.979 in the largest $(100, 25)$. This monotonic gain is consistent with the idea that the method benefits from longer time series. At the same time, for any fixed $T$, expanding the candidate variable set increases the complexity of selection, leading to reduced accuracy.

p-Lasso performs moderately well in this setting, with MCC values increasing from 0.723 to 0.851 as panel size grows. i-Lasso is more sensitive to $T$: while it performs relatively poorly for short panels (e.g., 0.529 in $(25, 25)$), its accuracy increases markedly with time series length, closely tracking PCA-MTB in high-$T$ configurations.

When $\phi = 0$, i.e., when the average factor loading is zero and pooling is less appropriate, differences across methods become more pronounced. PCA-MTB remains stable and continues to dominate, with MCC rising from 0.878 to 0.979 across configurations. p-Lasso, by contrast, is heavily affected: its MCC values remain below 0.71 in all cases and drop as low as 0.403 in $(25, 100)$. i-Lasso is more resilient than p-Lasso, showing improved performance with increasing $T$, and attaining MCC levels close to those of PCA-MTB in large panels.

As the signal dimension increases to $r = 5$, the estimation task becomes more challenging due to the higher complexity of the underlying structure. Nonetheless, PCA-MTB maintains strong performance, with MCC rising from 0.683 in a small-$T$ setting with a large number of candidate factors – four times the number of time series observations $(25, 100)$ – to ap-proximately 0.970 in the largest configurations. The method continues to clearly outperform the other approaches. While both p-Lasso and i-Lasso improve with panel size, they remain substantially less accurate than PCA-MTB. i-Lasso outperforms p-Lasso in nearly all cases and comes close to PCA-MTB when $T$ is large. However, its gains are more modest compared to the lower-$r$ case, highlighting the increased difficulty of signal recovery in higher dimensions.

The contrast sharpens when $\phi = 0$. PCA-MTB remains stable, with MCC ranging from 0.702 to 0.944, while p-Lasso performs poorly – especially in short panels – and i-Lasso shows improvement but consistently underperforms relative to PCA-MTB, unless $T$ is large.

Overall, PCA-MTB exhibits accurate and stable performance across all designs considered. Its robustness to variations in signal dimensionality, loading structure, and sample size suggests strong reliability in high-dimensional settings characterized by complex dependence structures.

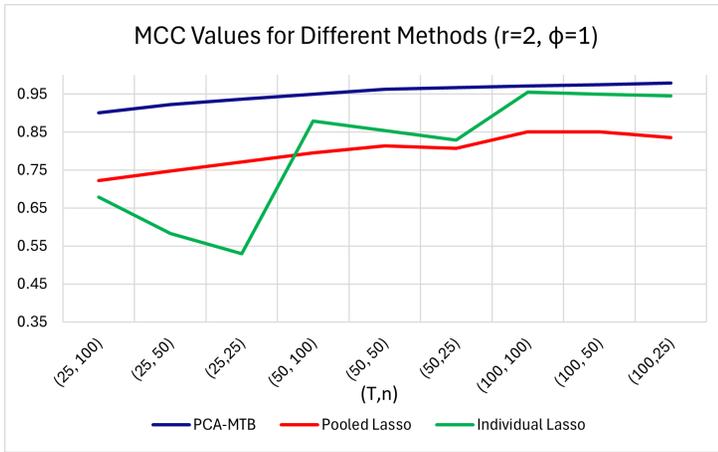

**MCC**: $r = 2$, $\phi = 1$

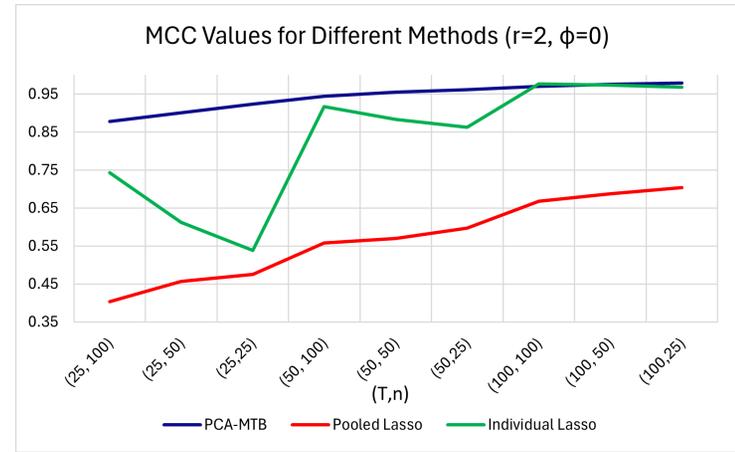

**MCC**: $r = 2$, $\phi = 0$

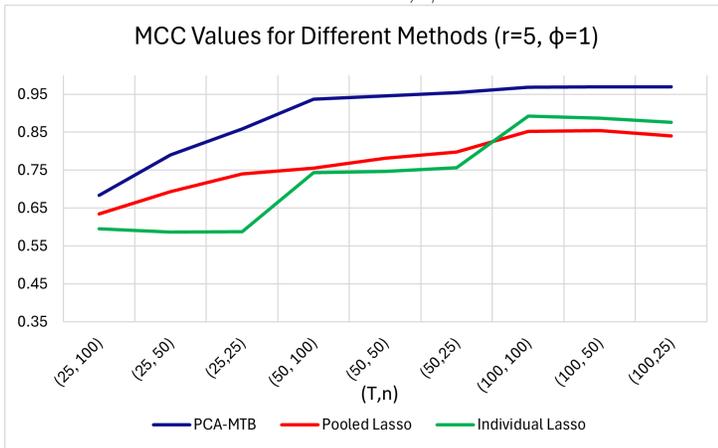

**MCC**: $r = 5$, $\phi = 1$

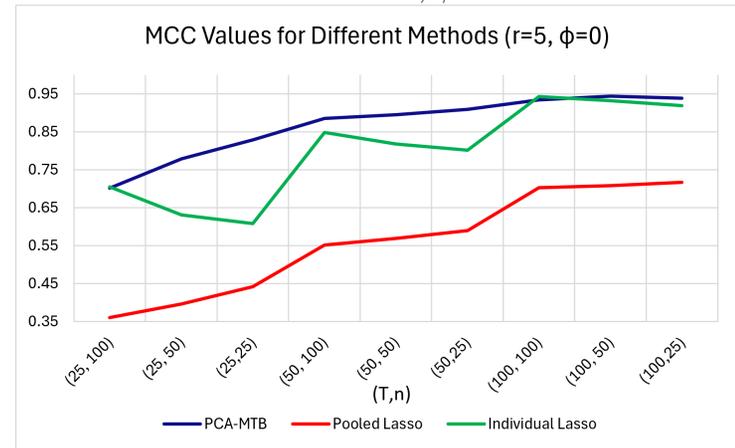

**MCC**: $r = 5$, $\phi = 0$

Figure C.1: MCC Performance Evaluation over different (T,N) values.

Table A5: MCC comparison across methods

|  |  | PCA-MTB | | | p-Lasso | | | i-Lasso | | |
|---|---|---|---|---|---|---|---|---|---|---|
| Design | T/n | 25 | 50 | 100 | 25 | 50 | 100 | 25 | 50 | 100 |
| $r=2, \phi=1$ | 25 | 0.937 | 0.923 | 0.901 | 0.772 | 0.747 | 0.723 | 0.530 | 0.584 | 0.679 |
|  | 50 | 0.967 | 0.964 | 0.951 | 0.808 | 0.814 | 0.796 | 0.830 | 0.854 | 0.879 |
|  | 100 | 0.979 | 0.976 | 0.972 | 0.836 | 0.851 | 0.852 | 0.946 | 0.950 | 0.956 |
| $r=2, \phi=0$ | 25 | 0.924 | 0.901 | 0.878 | 0.476 | 0.457 | 0.403 | 0.539 | 0.613 | 0.744 |
|  | 50 | 0.962 | 0.956 | 0.945 | 0.598 | 0.570 | 0.558 | 0.863 | 0.884 | 0.918 |
|  | 100 | 0.980 | 0.976 | 0.971 | 0.704 | 0.688 | 0.669 | 0.968 | 0.974 | 0.978 |
| $r=5, \phi=1$ | 25 | 0.859 | 0.791 | 0.684 | 0.741 | 0.693 | 0.635 | 0.588 | 0.587 | 0.596 |
|  | 50 | 0.955 | 0.946 | 0.937 | 0.798 | 0.782 | 0.756 | 0.757 | 0.747 | 0.744 |
|  | 100 | 0.970 | 0.970 | 0.969 | 0.841 | 0.854 | 0.853 | 0.876 | 0.887 | 0.893 |
| $r=5, \phi=0$ | 25 | 0.829 | 0.779 | 0.702 | 0.442 | 0.396 | 0.360 | 0.608 | 0.631 | 0.705 |
|  | 50 | 0.910 | 0.896 | 0.886 | 0.590 | 0.569 | 0.552 | 0.802 | 0.819 | 0.849 |
|  | 100 | 0.940 | 0.944 | 0.934 | 0.718 | 0.709 | 0.703 | 0.920 | 0.933 | 0.944 |

***Notes:*** Matthews Correlation Coefficient (MCC) is defined as

$$\text{MCC} = \frac{\text{TP} \cdot \text{TN} - \text{FP} \cdot \text{FN}}{\sqrt{(\text{TP} + \text{FP})(\text{TP} + \text{FN})(\text{TN} + \text{FP})(\text{TN} + \text{FN})}},$$

where TP, FP, TN, and FN denote the number of true positives, false positives, true negatives, and false negatives, respectively. MCC measures the correlation between predicted and true binary outcomes, reflecting how well the selected variables align with the actual ones. It ranges from $-1$ (perfect misclassification) to $+1$ (perfect selection), with 0 indicating random guessing.

Table A6: F1 comparison across methods

|  |  | PCA-MTB | | | p-Lasso | | | i-Lasso | | |
|---|---|---|---|---|---|---|---|---|---|---|
| Design | T/n | 25 | 50 | 100 | 25 | 50 | 100 | 25 | 50 | 100 |
| $r=2, \phi=1$ | 25 | 0.936 | 0.918 | 0.893 | 0.763 | 0.723 | 0.687 | 0.502 | 0.535 | 0.637 |
|  | 50 | 0.966 | 0.961 | 0.946 | 0.801 | 0.798 | 0.772 | 0.826 | 0.844 | 0.868 |
|  | 100 | 0.979 | 0.974 | 0.970 | 0.832 | 0.839 | 0.836 | 0.945 | 0.947 | 0.952 |
| $r=2, \phi=0$ | 25 | 0.924 | 0.896 | 0.872 | 0.476 | 0.450 | 0.390 | 0.512 | 0.569 | 0.713 |
|  | 50 | 0.961 | 0.953 | 0.940 | 0.596 | 0.559 | 0.544 | 0.861 | 0.877 | 0.911 |
|  | 100 | 0.979 | 0.974 | 0.968 | 0.699 | 0.676 | 0.653 | 0.968 | 0.972 | 0.976 |
| $r=5, \phi=1$ | 25 | 0.887 | 0.824 | 0.735 | 0.767 | 0.689 | 0.601 | 0.626 | 0.576 | 0.557 |
|  | 50 | 0.960 | 0.950 | 0.939 | 0.820 | 0.783 | 0.740 | 0.782 | 0.747 | 0.727 |
|  | 100 | 0.974 | 0.971 | 0.969 | 0.859 | 0.857 | 0.847 | 0.891 | 0.890 | 0.891 |
| $r=5, \phi=0$ | 25 | 0.862 | 0.810 | 0.746 | 0.461 | 0.397 | 0.348 | 0.644 | 0.623 | 0.683 |
|  | 50 | 0.919 | 0.901 | 0.891 | 0.611 | 0.567 | 0.536 | 0.823 | 0.821 | 0.843 |
|  | 100 | 0.946 | 0.946 | 0.934 | 0.739 | 0.709 | 0.692 | 0.930 | 0.935 | 0.943 |

***Notes:*** F1 is the harmonic mean of precision and recall, defined as

$$\text{F1} = \frac{2 \cdot \text{TP}}{2 \cdot \text{TP} + \text{FP} + \text{FN}},$$

where TP, FP, and FN denote the number of true positives, false positives, and false negatives, respectively. It balances the trade-off between discovery (recall) and accuracy (precision), and is especially informative under sparsity and class imbalance.

Table A7: Model size comparison across methods

|  |  | PCA-MTB | | | p-Lasso | | | i-Lasso | | |
|---|---|---|---|---|---|---|---|---|---|---|
| Design | T/n | 25 | 50 | 100 | 25 | 50 | 100 | 25 | 50 | 100 |
| $r=2, \phi=1$ | 25 | 2.322 | 2.427 | 2.600 | 3.578 | 3.933 | 4.359 | 6.361 | 5.904 | 4.655 |
|  | 50 | 2.172 | 2.204 | 2.286 | 3.220 | 3.276 | 3.505 | 2.969 | 2.864 | 2.722 |
|  | 100 | 2.109 | 2.133 | 2.157 | 3.001 | 2.956 | 2.991 | 2.282 | 2.273 | 2.246 |
| $r=2, \phi=0$ | 25 | 2.250 | 2.407 | 2.548 | 1.735 | 1.745 | 1.633 | 6.204 | 5.437 | 3.889 |
|  | 50 | 2.179 | 2.215 | 2.293 | 2.024 | 1.918 | 1.886 | 2.750 | 2.666 | 2.467 |
|  | 100 | 2.106 | 2.132 | 2.162 | 2.200 | 2.086 | 1.947 | 2.163 | 2.142 | 2.123 |
| $r=5, \phi=1$ | 25 | 5.722 | 5.628 | 5.516 | 8.297 | 9.963 | 12.406 | 11.209 | 12.727 | 13.514 |
|  | 50 | 5.412 | 5.511 | 5.617 | 7.385 | 8.031 | 8.892 | 7.948 | 8.623 | 9.051 |
|  | 100 | 5.095 | 5.118 | 5.142 | 6.780 | 6.819 | 6.981 | 6.308 | 6.319 | 6.317 |
| $r=5, \phi=0$ | 25 | 5.223 | 5.281 | 5.239 | 3.177 | 3.146 | 3.131 | 10.741 | 11.347 | 9.968 |
|  | 50 | 4.952 | 4.978 | 5.021 | 4.063 | 3.864 | 3.918 | 7.272 | 7.324 | 7.018 |
|  | 100 | 4.837 | 4.852 | 4.818 | 4.771 | 4.477 | 4.370 | 5.818 | 5.748 | 5.657 |

***Notes:*** Model size refers to the number of predictors selected by each method in a given replication. Values shown are the mean model sizes across Monte Carlo iterations.

Table A8: TDR comparison across methods

|  |  | PCA-MTB | | | p-Lasso | | | i-Lasso | | |
| --- | --- | --- | --- | --- | --- | --- | --- | --- | --- | --- |
| Design | T/n | 25 | 50 | 100 | 25 | 50 | 100 | 25 | 50 | 100 |
| | 25 | 0.900 | 0.874 | 0.835 | 0.651 | 0.600 | 0.558 | 0.344 | 0.378 | 0.488 |
| $r=2, \phi=1$ | 50 | 0.946 | 0.937 | 0.913 | 0.698 | 0.696 | 0.664 | 0.725 | 0.753 | 0.790 |
| | 100 | 0.965 | 0.957 | 0.950 | 0.741 | 0.753 | 0.749 | 0.909 | 0.913 | 0.921 |
| | 25 | 0.902 | 0.865 | 0.824 | 0.725 | 0.703 | 0.685 | 0.354 | 0.413 | 0.579 |
| $r=2, \phi=0$ | 50 | 0.941 | 0.929 | 0.908 | 0.756 | 0.755 | 0.746 | 0.777 | 0.802 | 0.855 |
| | 100 | 0.966 | 0.958 | 0.949 | 0.789 | 0.792 | 0.810 | 0.947 | 0.954 | 0.960 |
| | 25 | 0.833 | 0.767 | 0.664 | 0.635 | 0.539 | 0.443 | 0.460 | 0.410 | 0.393 |
| $r=5, \phi=1$ | 50 | 0.933 | 0.915 | 0.898 | 0.707 | 0.656 | 0.601 | 0.651 | 0.605 | 0.582 |
| | 100 | 0.970 | 0.965 | 0.961 | 0.763 | 0.762 | 0.747 | 0.812 | 0.811 | 0.812 |
| | 25 | 0.846 | 0.784 | 0.705 | 0.764 | 0.705 | 0.673 | 0.480 | 0.458 | 0.529 |
| $r=5, \phi=0$ | 50 | 0.937 | 0.917 | 0.900 | 0.794 | 0.783 | 0.762 | 0.708 | 0.706 | 0.740 |
| | 100 | 0.969 | 0.968 | 0.960 | 0.822 | 0.830 | 0.826 | 0.876 | 0.886 | 0.899 |

***Notes:*** The True Discovery Rate (TDR), also known as precision, is defined as

$$\text{TDR} = \frac{\text{TP}}{\text{TP}+\text{FP}},$$

where TP and FP denote the number of true and false positives, respectively. TDR reflects the proportion of selected variables that are truly relevant. Higher values indicate greater accuracy in identifying true signals while avoiding false discoveries.

Table A9: FDR comparison across methods

|  |  | PCA-MTB | | | p-Lasso | | | i-Lasso | | |
| --- | --- | --- | --- | --- | --- | --- | --- | --- | --- | --- |
| Design | T/n | 25 | 50 | 100 | 25 | 50 | 100 | 25 | 50 | 100 |
| | 25 | 0.100 | 0.126 | 0.165 | 0.349 | 0.400 | 0.442 | 0.656 | 0.622 | 0.512 |
| $r=2, \phi=1$ | 50 | 0.054 | 0.063 | 0.087 | 0.302 | 0.304 | 0.336 | 0.275 | 0.247 | 0.210 |
| | 100 | 0.035 | 0.043 | 0.050 | 0.259 | 0.247 | 0.251 | 0.091 | 0.087 | 0.079 |
| | 25 | 0.098 | 0.135 | 0.176 | 0.275 | 0.297 | 0.315 | 0.646 | 0.587 | 0.421 |
| $r=2, \phi=0$ | 50 | 0.059 | 0.071 | 0.092 | 0.244 | 0.245 | 0.254 | 0.223 | 0.198 | 0.145 |
| | 100 | 0.034 | 0.042 | 0.051 | 0.211 | 0.208 | 0.190 | 0.053 | 0.046 | 0.040 |
| | 25 | 0.167 | 0.233 | 0.336 | 0.365 | 0.461 | 0.557 | 0.540 | 0.590 | 0.607 |
| $r=5, \phi=1$ | 50 | 0.067 | 0.085 | 0.102 | 0.293 | 0.344 | 0.399 | 0.349 | 0.395 | 0.418 |
| | 100 | 0.030 | 0.035 | 0.039 | 0.237 | 0.238 | 0.253 | 0.188 | 0.189 | 0.188 |
| | 25 | 0.154 | 0.216 | 0.295 | 0.236 | 0.295 | 0.327 | 0.520 | 0.542 | 0.471 |
| $r=5, \phi=0$ | 50 | 0.063 | 0.083 | 0.100 | 0.206 | 0.217 | 0.238 | 0.292 | 0.294 | 0.260 |
| | 100 | 0.031 | 0.032 | 0.040 | 0.178 | 0.170 | 0.174 | 0.124 | 0.114 | 0.101 |

***Notes:*** The False Discovery Rate (FDR) is defined as

$$\text{FDR} = \frac{\text{FP}}{\text{TP}+\text{FP}},$$

where FP and TP denote the number of false and true positives, respectively. FDR reflects the proportion of selected predictors that are in fact irrelevant. Lower values indicate greater reliability in variable selection.

Table A10: TPR comparison across methods

|  |  | PCA-MTB | | | p-Lasso | | | i-Lasso | | |
|---|---|---|---|---|---|---|---|---|---|---|
| Design | T/n | 25 | 50 | 100 | 25 | 50 | 100 | 25 | 50 | 100 |
| $r=2, \phi=1$ | 25 | 0.996 | 0.995 | 0.995 | 1.000 | 1.000 | 1.000 | 1.000 | 1.000 | 1.000 |
|  | 50 | 1.000 | 1.000 | 1.000 | 1.000 | 1.000 | 1.000 | 1.000 | 1.000 | 1.000 |
|  | 100 | 1.000 | 1.000 | 1.000 | 1.000 | 1.000 | 1.000 | 1.000 | 1.000 | 1.000 |
| $r=2, \phi=0$ | 25 | 0.970 | 0.966 | 0.962 | 0.520 | 0.495 | 0.433 | 1.000 | 1.000 | 1.000 |
|  | 50 | 0.995 | 0.994 | 0.995 | 0.652 | 0.606 | 0.589 | 1.000 | 1.000 | 1.000 |
|  | 100 | 1.000 | 1.000 | 1.000 | 0.763 | 0.726 | 0.688 | 1.000 | 1.000 | 1.000 |
| $r=5, \phi=1$ | 25 | 0.943 | 0.867 | 0.755 | 1.000 | 1.000 | 0.999 | 1.000 | 1.000 | 1.000 |
|  | 50 | 0.996 | 0.993 | 0.990 | 1.000 | 1.000 | 1.000 | 1.000 | 1.000 | 1.000 |
|  | 100 | 0.982 | 0.981 | 0.982 | 1.000 | 1.000 | 1.000 | 1.000 | 1.000 | 1.000 |
| $r=5, \phi=0$ | 25 | 0.878 | 0.827 | 0.749 | 0.437 | 0.382 | 0.338 | 1.000 | 1.000 | 1.000 |
|  | 50 | 0.917 | 0.901 | 0.891 | 0.599 | 0.548 | 0.524 | 1.000 | 1.000 | 1.000 |
|  | 100 | 0.932 | 0.934 | 0.919 | 0.743 | 0.697 | 0.677 | 1.000 | 1.000 | 1.000 |

***Notes:*** The True Positive Rate (TPR), also known as recall or sensitivity, is defined as

$$\text{TPR} = \frac{\text{TP}}{\text{TP} + \text{FN}},$$

where TP and FN denote the number of true positives and false negatives, respectively. TPR measures the proportion of relevant variables correctly identified. A TPR of 1 indicates perfect recovery of all true signals.

Table A11: FPR comparison across methods

|  |  | PCA-MTB | | | p-Lasso | | | i-Lasso | | |
|---|---|---|---|---|---|---|---|---|---|---|
| Design | T/n | 25 | 50 | 100 | 25 | 50 | 100 | 25 | 50 | 100 |
| $r=2, \phi=1$ | 25 | 0.013 | 0.009 | 0.006 | 0.063 | 0.039 | 0.024 | 0.162 | 0.075 | 0.026 |
|  | 50 | 0.007 | 0.004 | 0.003 | 0.049 | 0.026 | 0.015 | 0.036 | 0.017 | 0.007 |
|  | 100 | 0.004 | 0.003 | 0.002 | 0.040 | 0.019 | 0.010 | 0.010 | 0.005 | 0.002 |
| $r=2, \phi=0$ | 25 | 0.012 | 0.009 | 0.006 | 0.028 | 0.015 | 0.008 | 0.156 | 0.066 | 0.019 |
|  | 50 | 0.008 | 0.005 | 0.003 | 0.029 | 0.014 | 0.007 | 0.028 | 0.013 | 0.005 |
|  | 100 | 0.004 | 0.003 | 0.002 | 0.027 | 0.013 | 0.006 | 0.006 | 0.003 | 0.001 |
| $r=5, \phi=1$ | 25 | 0.040 | 0.026 | 0.017 | 0.132 | 0.099 | 0.074 | 0.207 | 0.140 | 0.081 |
|  | 50 | 0.017 | 0.011 | 0.007 | 0.095 | 0.061 | 0.039 | 0.098 | 0.066 | 0.039 |
|  | 100 | 0.007 | 0.004 | 0.002 | 0.071 | 0.036 | 0.020 | 0.044 | 0.024 | 0.013 |
| $r=5, \phi=0$ | 25 | 0.033 | 0.023 | 0.015 | 0.040 | 0.025 | 0.014 | 0.191 | 0.115 | 0.047 |
|  | 50 | 0.015 | 0.009 | 0.006 | 0.043 | 0.023 | 0.013 | 0.076 | 0.042 | 0.019 |
|  | 100 | 0.007 | 0.004 | 0.002 | 0.042 | 0.020 | 0.010 | 0.027 | 0.014 | 0.006 |

***Notes:*** The False Positive Rate (FPR) is defined as

$$\text{FPR} = \frac{\text{FP}}{\text{FP} + \text{TN}},$$

where FP and TN denote the number of false positives and true negatives, respectively. FPR measures the proportion of irrelevant variables that were incorrectly selected. A lower FPR indicates fewer false discoveries.